\PassOptionsToPackage{square,numbers,comma,sort&compress,super}{natbib} 
\documentclass[aps,prl,twocolumn,a4paper,superscriptaddress]{revtex4-2}

\usepackage[paperwidth=215mm,paperheight=300mm,centering,hmargin=1.6cm,vmargin=2cm]{geometry}
\usepackage{graphicx}
\usepackage{amsmath}

\usepackage{latexsym,stmaryrd}
\usepackage[dvipsnames]{xcolor}
\usepackage[normalem]{ulem}
\usepackage{float}

\usepackage{siunitx} 
\usepackage{physics} 
\usepackage{chemformula} 
\usepackage{titlesec}
\usepackage{gensymb}

\titleformat{\section}[hang]{\bfseries}{}{1em}{}
\titlespacing*{\section}{0pt}{2.5ex plus 1ex minus .2ex}{0.2ex plus .2ex}
\titleformat{\subsection}[hang]{\bfseries}{}{1em}{}
\titlespacing*{\subsection}{0pt}{2.5ex plus 1ex minus .2ex}{0.2ex plus .2ex}

\renewcommand{\figurename}{\textbf{Fig.}}

\makeatletter
\renewcommand*{\fnum@figure}{{\normalfont\bfseries \figurename~\thefigure}}
\makeatother

\begin{document}

\title{Self-assembly of atomic-scale photonic cavities}

\author{Ali Nawaz Babar}
\email{anaba@dtu.dk}
\affiliation{Department of Electrical and Photonics Engineering, DTU Electro, Technical University of Denmark, Building 343, DK-2800 Kgs. Lyngby, Denmark}
\affiliation{NanoPhoton - Center for Nanophotonics, Technical University of Denmark, Ørsteds Plads 345A, DK-2800 Kgs. Lyngby, Denmark.}

\author{Thor August Schimmell Weis}
\affiliation{Department of Electrical and Photonics Engineering, DTU Electro, Technical University of Denmark, Building 343, DK-2800 Kgs. Lyngby, Denmark}

\author{Konstantinos Tsoukalas}
\affiliation{Department of Electrical and Photonics Engineering, DTU Electro, Technical University of Denmark, Building 343, DK-2800 Kgs. Lyngby, Denmark}

\author{Shima Kadkhodazadeh}
\affiliation{NanoPhoton - Center for Nanophotonics, Technical University of Denmark, Ørsteds Plads 345A, DK-2800 Kgs. Lyngby, Denmark.}
\affiliation{DTU Nanolab, Technical University of Denmark, Building 307,
DK-2800 Kgs. Lyngby, Denmark}

\author{Guillermo Arregui}
\affiliation{Department of Electrical and Photonics Engineering, DTU Electro, Technical University of Denmark, Building 343, DK-2800 Kgs. Lyngby, Denmark}

\author{Babak Vosoughi Lahĳani}
\affiliation{Department of Electrical and Photonics Engineering, DTU Electro, Technical University of Denmark, Building 343, DK-2800 Kgs. Lyngby, Denmark}
\affiliation{NanoPhoton - Center for Nanophotonics, Technical University of Denmark, Ørsteds Plads 345A, DK-2800 Kgs. Lyngby, Denmark.}

\author{Søren Stobbe }
\email{ssto@dtu.dk}
\affiliation{Department of Electrical and Photonics Engineering, DTU Electro, Technical University of Denmark, Building 343, DK-2800 Kgs. Lyngby, Denmark}
\affiliation{NanoPhoton - Center for Nanophotonics, Technical University of Denmark, Ørsteds Plads 345A, DK-2800 Kgs. Lyngby, Denmark.}

\homepage{}

\date{\today}
\maketitle

\small

\bfseries\noindent
Despite tremendous progress in the research on self-assembled nanotechnological building blocks such as macromolecules~\cite{winfree1998design}, nanowires~\cite{heiss2013self}, and two-dimensional materials~\cite{sun2014generalized}, synthetic self-assembly methods bridging nanoscopic to macroscopic dimensions 
remain unscalable and inferior to biological self-assembly.
In contrast, planar semiconductor technology has had an immense technological impact owing to its inherent scalability, yet it appears unable to reach the atomic dimensions enabled by self-assembly.
Here we use surface forces including Casimir-van der Waals interactions~\cite{Casimir_realmaterials} to deterministically self-assemble and self-align suspended silicon nanostructures with void features well below the length scales possible with conventional lithography and etching \cite{Marcus, IRDS_lithography_2022}, despite using nothing more than conventional lithography and etching. The method is remarkably robust and the threshold for self-assembly depends monotonically on all governing parameters across thousands of measured devices. 
We illustrate the potential of these concepts by fabricating nanostructures, which are impossible to make with any other known method: Waveguide-coupled high-$Q$ silicon photonic cavities~\cite{Koenderink_Shrinkinglight, Lodahl} that confine telecom photons to 2 nm air gaps with an aspect ratio of 100, corresponding to mode volumes more than 100 times below the diffraction limit.
Scanning transmission electron microscopy measurements confirm the ability to build devices even with subnanometer dimensions. Our work constitutes the first steps towards a new generation of fabrication technology that combines the atomic dimensions enabled by self-assembly with the scalability of planar semiconductors.\\

\normalfont

\begin{figure*}[t]
\centering
 \includegraphics[width=\textwidth]{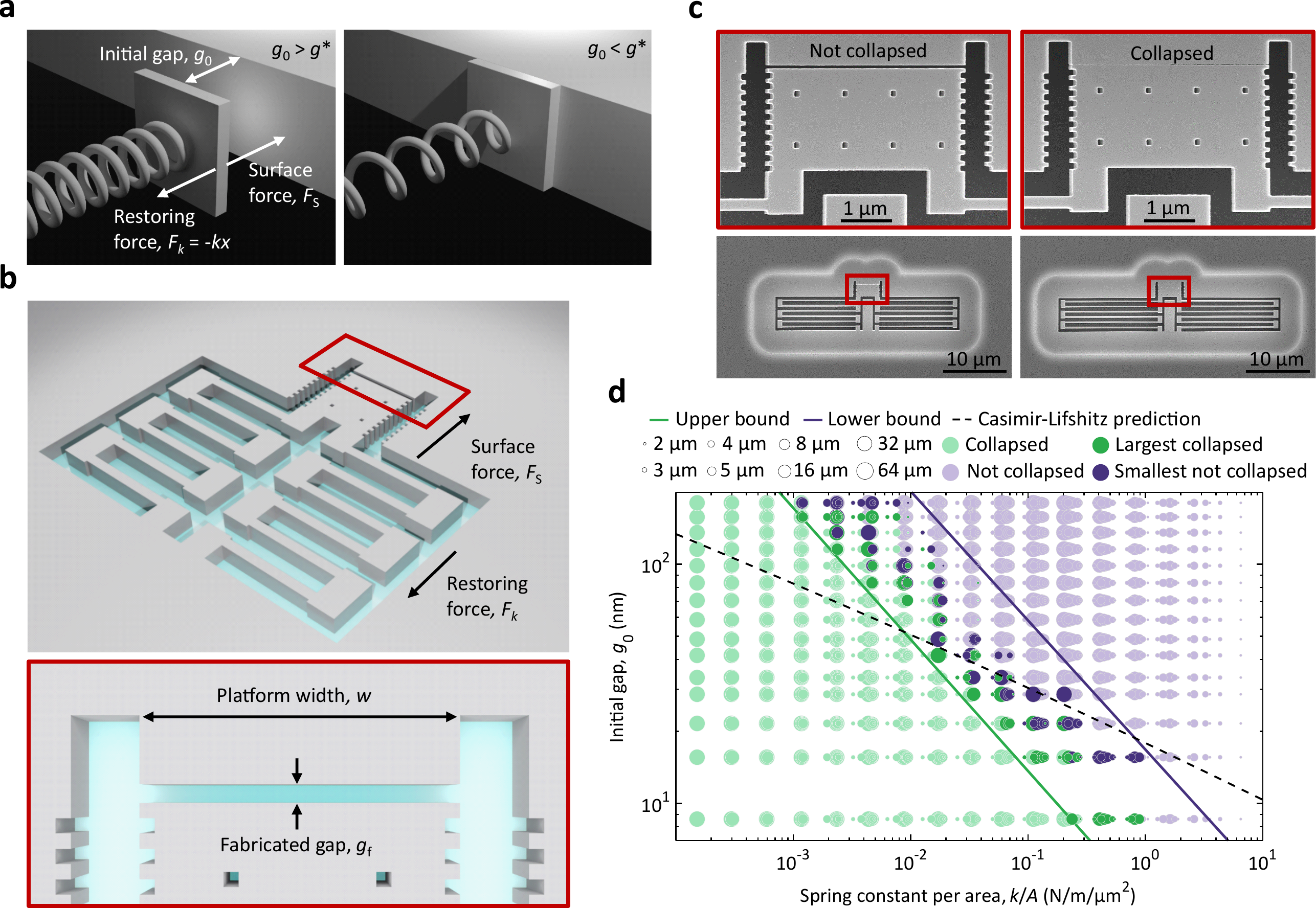}
    \caption{\textbf{Deterministic in-plane self-assembly of suspended silicon platforms by surface forces.} \textbf{a}, Experimental concept for mapping the design space for self-assembly exploiting the pull-in instabilities associated with on-chip surface forces. The balance between the nonlinear surface forces and linear spring forces means that devices with initial gaps, $g_0$, that are larger (smaller) than the critical gap, $g^*$, do not collapse (collapse deterministically) as shown on the left (right). \textbf{b}, Illustration of the realization of the experiment in a silicon-on-insulator platform before release by underetching the silicon device layer. The force balance depends on the platform width, $w$, the initial gap, $g_\text{0}$, and the spring constant, $k$. After the release etch, the compressively strained silicon device layer expands and reaches a new equilibrium position relative to the fabricated gap, $g_\text{f}$, which determines the initial gap, i.e., $g_\text{0} = g_\text{f} -19.4$~nm. \textbf{c}, Tilted-view ($20 ^{\circ}$) SEM images of two devices after the membrane release by removal of the sacrificial buried oxide layer with the same platform width of \SI{4}{\um}, initial gap of \SI{41}{\nm} but with different spring constants of 0.038 N/m (left, not collapsed) and 0.019 N/m (right, deterministically collapsed). \textbf{d}, Measured map of the design space for self-assembly with compliant silicon structures, obtained by characterizing 1536 platforms by SEM. The green-filled circles represent the collapsed platforms, and the purple-filled represent non-collapsed platforms. The different sizes of the circles represent the different widths of the platforms. The dark purple circle indicate the smallest initial gap not leading to a collapsed platform and the dark green circle indicates the largest initial gap leading to a collapsed platform. All the devices below the upper bound collapse, while those above the lower bound do not. \ }
    \label{fig:1}
\end{figure*}

The fabrication of functional materials and devices at the micro- and nanoscale typically follows either a top-down approach, which uses complex sequences of planar technology such as lithography and etching, or a bottom-up approach, where structures are self-assembled using various effects, such as van der Waals, electrostatic, capillary, or hydrogen-bonding forces~\cite{TopdownmeetsBottomup, InterparticleandExternalforces, Nanoparticleselfassembly}. While top-down nanofabrication underpins the unique scalability of semiconductor technology, the bottom-up approach has enabled a wide range of research on devices with near-atomic dimensions. Such miniaturization is crucial for a wealth of research and technology that rely on an increased surface-to-volume ratio, strong field gradients, or quantum effects. Examples include ultrahigh-frequency surface-acoustic-wave resonators~\cite{wang_high_2018}, superconducting nanowire single-photon detectors~\cite{marsili_single-photon_2011}, X-ray zone plates~\cite{chao_soft_2005}, and nanopore sequencing of DNA strands~\cite{deamer_three_2016}. In addition, the vision of complex and often hybrid and hetero-integrated devices relying on technology at the few-nanometer scale, sometimes denoted More than Moore, is now central to a wide range of research ranging from biosensing~\cite{liu_sculpting_2018} to quantum technologies~\cite{QuantumOptics}. 
However, the miniaturization of semiconductor technology has slowed to the point where the so-called technology nodes no longer indicate physical dimensions. For example, the current industry roadmap~\cite{IRDS_lithography_2022} forecasts no lateral lithography features (minimum half-pitch or physical gate length) below 8 nm for the next 15 years. However, state-of-the-art today is already denoted the ``3 nm node"~\cite{liu2021_transistors_nodes}. At the same time, while bottom-up approaches can achieve feature sizes down to atomic scales, synthetic self-assembly remains far from capable of replicating the hierarchical and scalable self-assembly in biological systems~\cite{Nanoparticleselfassembly, Guidedselfassembly,zhang_fabrication_2003}. A practical consequence is that a wealth of research on bottom-up nanotechnology for information technology always had to rely on top-down technology for the interconnect architecture, e.g., lithographically defined wires or waveguides are needed to contact single-molecule devices~\cite{thiele_electrically_2014,zhang_gdc82_2020} or single-quantum-dot devices~\cite{Lodahl}.
Combining the scalability of top-down planar technology with the resolution of bottom-up approaches would open vast perspectives for both research and technology \cite{TopdownmeetsBottomup}, but they are commonly considered disjoint. Strategies for combining them are scarce~\cite{luo2022high_nanogaps_selfassembly, F.Nealey.selfassembly} 
and a pathway for their direct integration was so far missing. 

Recent developments have brought miniaturization to the center stage also in photonics. Increasing the strength of the interaction between light and matter has been a central goal in quantum optics and photonics for decades and traditionally followed either of two paths: dielectric nanocavities offering high quality factors and scalable waveguide integration but limited confinement or, alternatively, metal nanocavities offering strong spatial confinement due to plasmonic effects~\cite{duan_nanoplasmonics_2012,liu_sculpting_2018,kim_capillary-force-induced_2020} but suffering from absorption and limited quality factors. The two approaches can be combined in a hybrid cavity-antenna system~\cite{palstra_hybrid_2019}, but plasmon resonances limit their applicability to visible and short infrared wavelengths. Using dielectric bowtie cavities, it is possible to combine strong spatial confinement with high quality factors at telecom wavelengths, which was predicted in 2005~\cite{Gondarenko_spontaneous_periodicpattern_PRL_2006} but demonstrated only very recently~\cite{Marcus}, in part because realistic designs were missing and in part because experimental progress was impeded by the extreme requirements posed on the nanofabrication. Dielectric bowtie cavities harness the field discontinuities at material boundaries to strongly confine light inside dielectrics~\cite{lightningrod}, and hold the promise of unprecedented light-matter interaction strengths, fostering new developments in nanolasers and optical interconnects~\cite{nanolaser, EDClaser}, nonlinear photonics~\cite{Choi}, all-optical switching~\cite{switching}, cavity quantum electrodynamics \cite{Lodahl}, and cavity optomechanics~\cite{bozkurt_deep_2022}. In addition, dielectric bowtie cavities may also enable probing fundamental limits to the light-matter interaction strength~\cite{Chao_physical_limits_2022,panuski_fundamental_2020} as well as investigating the validity of the continuum model of electromagnetism, which is known to break down in plasmonics~\cite{Mortensen_plasmonic_nanostructures_NatCom_2014} but has only recently been explored in dielectrics~\cite{bharadwaj2022pico}. Since the width of the bowtie determines the electromagnetic field enhancement~\cite{lightningrod}, this has put progress in planar semiconductor nanofabrication at the forefront of research in nanophotonics. The first experiment~\cite{Marcus} demonstrating confinement of light below the diffraction limit in dielectric bowtie cavities employed 8 nm wide silicon bridges with an aspect ratio of 30 and although minor improvements along this route may be possible, it appears futile to try to scale conventional lithography and etching to atomic dimensions with aspect ratios exceeding 100. Void or low-refractive-index features with extreme aspect ratios are especially challenging to fabricate, but they are required for some of the most radical applications of nanocavities, such as bulk nonlinearities operating at the single-photon level~\cite{Choi} and single-photon emitters for quantum photonic integrated circuits~\cite{hollenbach2022wafer}. 

Here we propose and demonstrate a novel approach to the manufacturing of semiconductor devices with unprecedented dimensions, namely using the ubiquitous surface forces that act on objects separated by a few tens of nanometers, i.e., the van der Waals force and the Casimir force~\cite{Casimir_realmaterials}.
The two are different limits of the same force that arises due to the fluctuations of the quantum vacuum~\cite{Casimir_realmaterials} and are normally considered nuisances that cause device failure of micro- and nanomechanical devices~\cite{Roukes_stiction_PRB_2001}. In contrast, our experiments aim to harness these forces to enable controlled, deterministic, and directional collapses to fabricate nanostructures with atomic-scale dimensions.

\section*{Deterministic self-assembly by surface forces}
Self-assembly is possible when components are free to move and can adhere to each other, which is often realized in liquid environments~\cite{Battulga}, but our method takes place in a gas or vacuum. To prevent the components from falling onto the substrate due to gravitational forces while still being free to move, we suspend and attach the components to the surrounding frame by springs, which are etched out of silicon. Common to the surface forces are their power-law dependence on the gap between objects, $g$, leading to pull-in instabilities as illustrated in Fig.~\ref{fig:1}a: When the nonlinear attractive surface force, $F_{\text{s}}$, overwhelms the opposing linear restoring force, $F_{k}$, at a critical gap~\cite{palasantzas2022}, $g^*$, the pull-in instability occurs. The suspended components collapse deterministically and subsequently adhere to each other by van der Waals forces, resulting in a structurally stable self-assembled device.

Although the surface forces are well understood theoretically \cite{Casimir_realmaterials, Rodriguez}, their exact numerical values are difficult to determine because they depend strongly on parameters such as surface treatment, doping level, and fabrication imperfections~\cite{burger2020comparison}. Therefore, the starting point of our investigation is to map the surface-force instability as a function of geometry, thus providing design rules for self-assembly by directed collapses. This experiment is implemented in a silicon-on-insulator platform as illustrated in Fig.~\ref{fig:1}b, using suspended silicon platforms in close proximity to a rigid and anchored silicon structure. The platforms are attached to the frame by two symmetric folded cantilever springs of spring constant $k$, separated from the anchored part by a gap, $g_\text{f}$. Our devices are defined using electron-beam lithography and reactive-ion etching.
Subsequently, the platforms are released from the  substrate by selective underetching of the oxide layer using anhydrous vapor-phase hydrofluoric acid. The released platforms will collapse or not collapse in-plane onto the anchored structure depending on whether the initial gap, $g_\text{0}$, is larger than the critical gap, $g^*$, which in turn depends on $k$ and $w$. This is illustrated in Fig.~\ref{fig:1}c, which shows representative scanning electron microscope (SEM) images of two devices with the same $w$ and $g_\text{0}$, but different $k$. The stiffer spring provides enough restoring force for the platform to reach a stable equilibrium at a small displacement without collapsing, while that with a smaller spring constant does not, leading to a deterministic and directed collapse.

\begin{figure*}[t]
\centering
 \includegraphics[width=\textwidth]{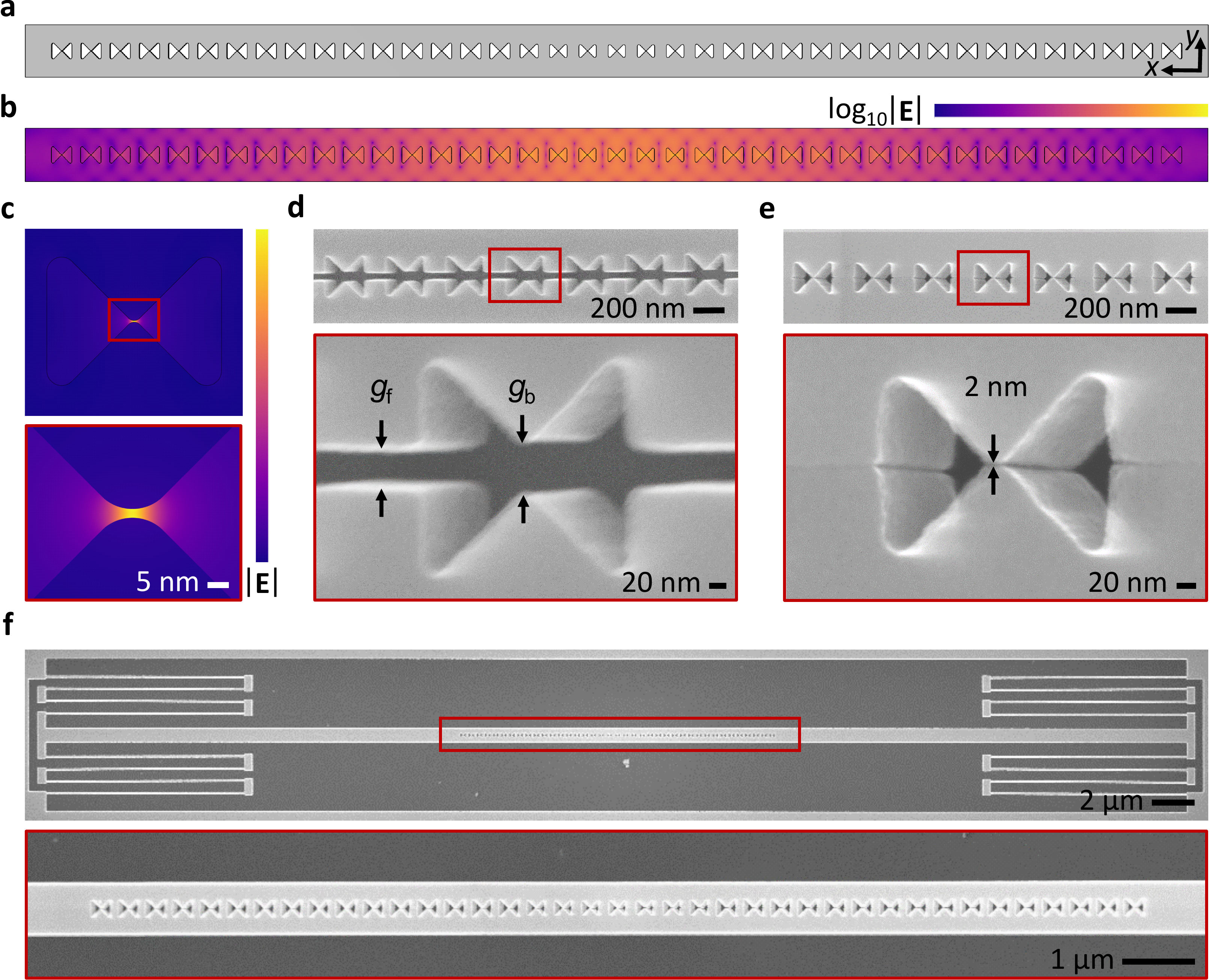}
    \caption{\textbf{Design and fabrication of a self-assembled silicon nanobeam bowtie cavity.} \textbf{a}, The geometry of the nanobeam bowtie cavity. \textbf{b}, Normalized electric field of the cavity mode in log-scale, $\log_{10}|\mathbf{E}|$. \textbf{c}, Normalized electric field, $|\mathbf{E}|$, of the central bowtie unit cell of the nanobeam cavity, showing that light is confined to a 2 nm air gap. \textbf{d}, Tilted ($20 ^{\circ}$) SEM image of the central part of the cavity before the release etch triggers the self-assembly of the two parts initially separated by $g_\text{f}$ = 50 nm, except at the bowtie where the distance is $g_\text{b}$. \textbf{e}, Tilted ($20 ^{\circ}$) SEM of the central part of a nanobeam cavity after self-assembly, with the approximately 2 nm gap indicated in the zoom-in. \textbf{f}, Top-view SEM image of the full device, including the spring suspension.}
    \label{fig:2}
\end{figure*}

\begin{figure*}[t]
\centering
 \includegraphics[width=\textwidth]{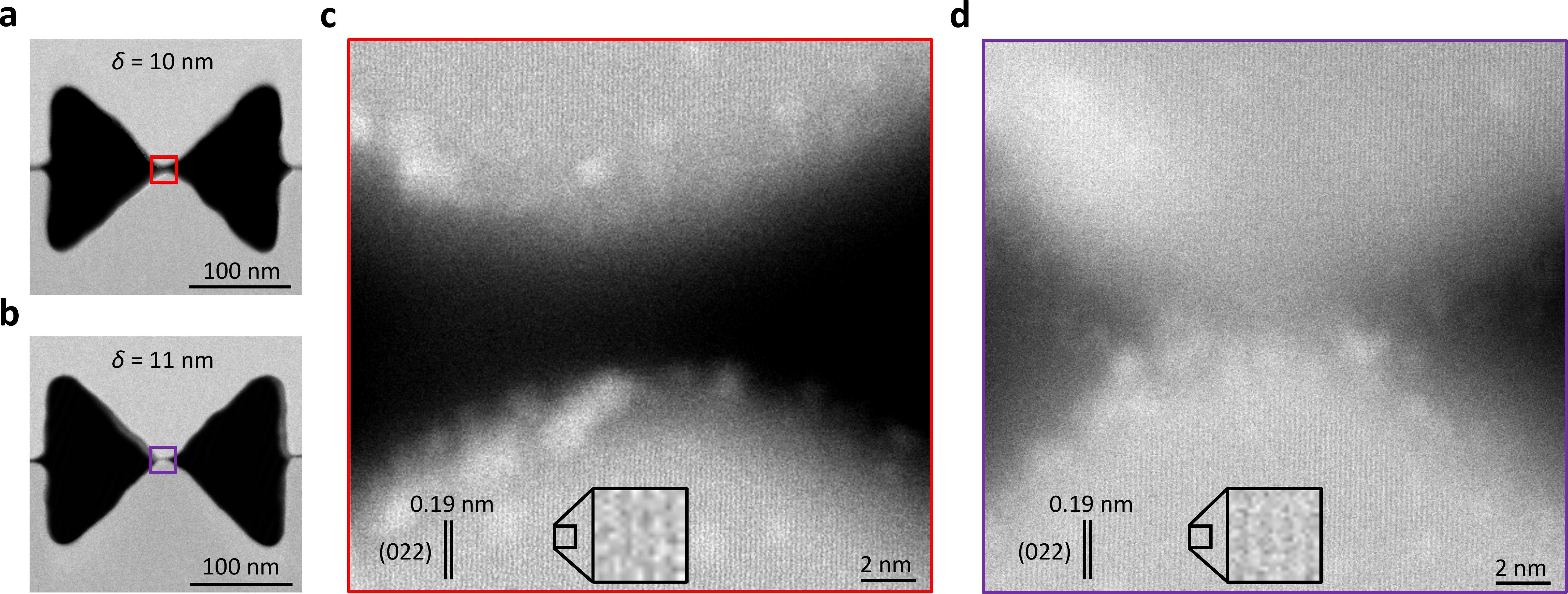}
    \caption{\textbf{High-resolution imaging of self-assembled air bowties.} \textbf{a-b}, Top-view scanning transmission electron microscope (STEM) image of the central bowtie unit cell of a self-assembled nanobeam cavity fabricated with offsets $\delta$ = 10 nm and $\delta$ = 11 nm, where $\delta$ represents the relative distance between the bowtie tip and half-nanobeam flat edge on the lithographic mask. \textbf{c-d}, High-resolution STEM image of the bowtie tip region highlighted with a box in (\textbf{a-b}) for $\delta$ = 10 nm and $\delta$ = 11 nm. The lower part of the structure is tilted to the [100] zone axis, revealing the (022) planes of the silicon crystal with an interplanar distance of 0.19 nm, which we use for accurately calibrating the scale bar.}
    \label{fig:3}
\end{figure*}

To map out the set of geometries that lead to directed in-plane collapses, we fabricate 2688 devices distributed across two samples (Sample A: 1536 devices; Sample B: 1152 devices) with varying values of $w$, $g_\text{f}$, and $k$ and here we discuss only Sample A (see Methods for details on parameters and data for Sample B, which reproduce the results from Sample A). Note that $g_\text{f}$ differs from the initial gap, $g_\text{0}$, because of a 19.4 nm displacement due to the release of the built-in compressive thermal stress after underetching, i.e., $g_\text{0} = g_\text{f} -19.4$~nm (see Supplementary Section S1.1 and S1.2 for details). We perform systematic SEM characterization of all devices after underetching and record which structures collapse and which do not. The resulting data is shown in Fig.~\ref{fig:1}d. For fixed values of $w$ and $k$, we identify two gaps: the largest value of $g_\text{0}$ for which the collapse occurs and the smallest value of $g_\text{0}$ for which the collapse does not occur. Using those gaps, we find that all platforms for which $g_0 < 3.8(k/A)^{-0.55}$ collapse and all platforms for which $g_0 > 16.8(k/A)^{-0.54}$ do not collapse. Given the significant sample size and yield of the experiment, i.e., only 11 devices out of 2688 devices failed due to out-of-plane collapse and/or lithographic errors, these thresholds provide the essential design rules for realizing suspended silicon devices with high-aspect-ratio gaps that avoid unintended pull-in instabilities, such as nano-opto-electromechanical systems~\cite{Midolo}, or, in the opposite limit, the criterion for deterministic self-assembly.

We include in Fig.~\ref{fig:1}d the critical gap calculated with the Lifshitz theory of the Casimir-van der Waals force in the proximity force approximation (PFA) for two silicon slabs (see Supplementary Section S1.3). We observe good agreement with the measured collapse threshold in the range where the PFA is expected to be valid, i.e., for gaps in the range of 20 to 50 nm. The model deviates from our experiment for small gaps, but systematic errors in SEM measurements of few-nanometer features can be very significant. The platforms are found to be more prone to collapse than predicted by the model for larger initial gaps and smaller spring constants, which indicates additional attractive contributions to the net surface force, such as electrostatic surface effects~\cite{behunin2012} and effects beyond the PFA~\cite{gies2006casimir}, which are both expected to be more important for large gaps. In any case, our static collapse experiment does not aim to replicate the abundance of accurate dynamical measurements of the Casimir force available in the literature~\cite{Casimir_realmaterials, Lamoreaux, Rodriguez, palasantzas2022, Roukes_stiction_PRB_2001}, but rather to map out the phase space separating collapse from non-collapse in a practical setting relevant for self-assembly. We note that the largest initial gap leading to collapse and the smallest gap not leading to collapse are adjacent data points across our entire data set of 2688 devices, which evidences a robust and reproducible method (see Supplementary Information Section 1.2 for the full data set and further discussions).

 \begin{figure*}[t]
\centering
 \includegraphics[width=\textwidth]{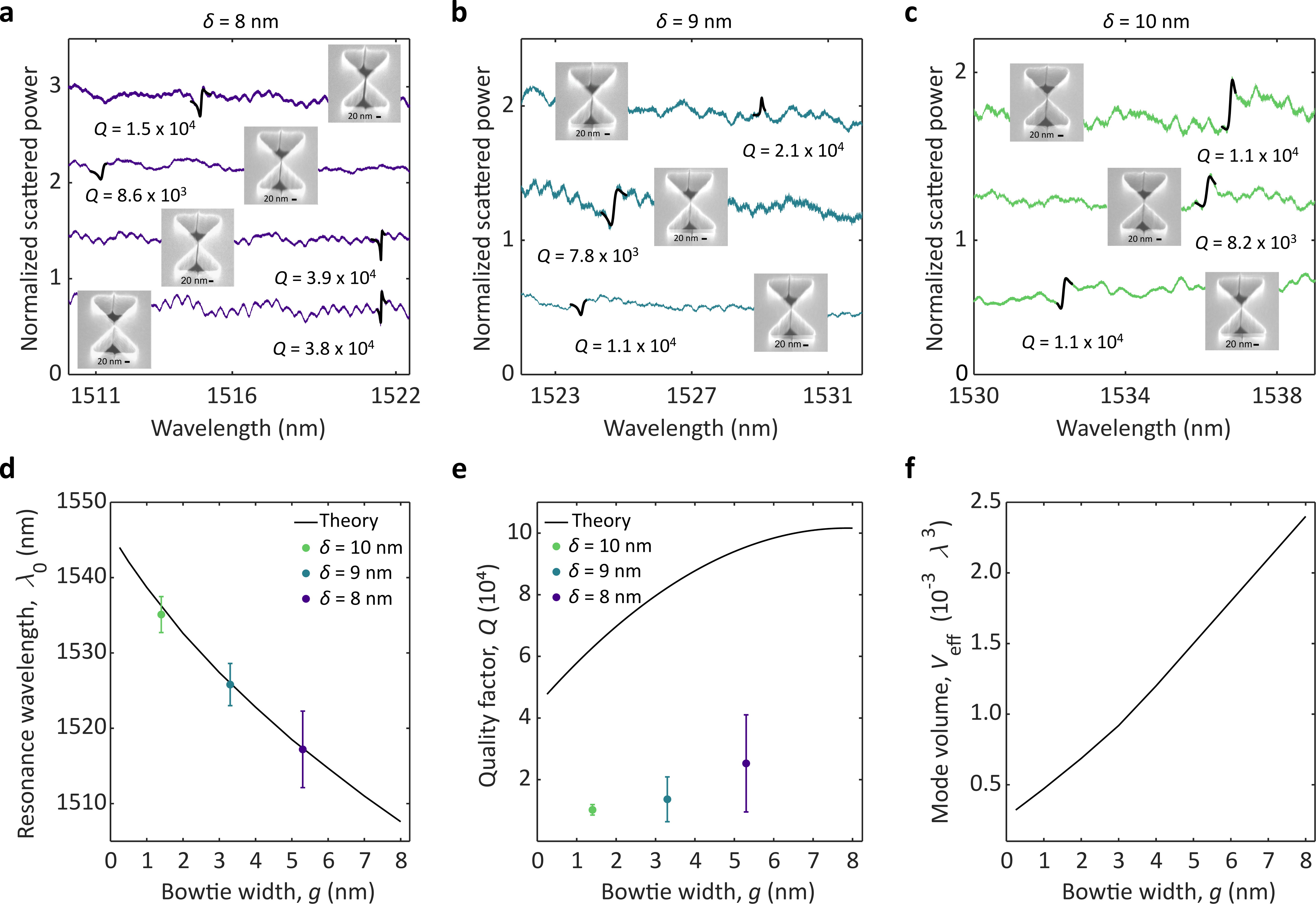}
    \caption{\textbf{Resonant scattering from self-assembled nanobeam cavities.} \textbf{a}-\textbf{c}, Normalized scattering spectra of three sets of nominally identical cavities with offsets $\delta$ = 8, 9, and 10 nm. On each panel, the spectra are shifted vertically, for clarity, and the resonant modes are fitted with Fano lineshapes (black lines). The extracted $Q$-factors and a tilted SEM of the central bowtie unit cell are given. \textbf{d}, Simulated and measured resonant wavelengths as a function of the bowtie width, $g$, where the simulated nanobeam uses all measured fabricated dimensions including a 2 nm native oxide layer and the experimental bowtie widths are estimated following Supplementary Section S3.2. The theory curve is shifted up by 15 nm. \textbf{e}, Calculated and measured quality factor versus bowtie width. The vertical error bars in (\textbf{d}) and (\textbf{e}) correspond to the standard deviation obtained from (\textbf{a}-\textbf{c}). \textbf{f}, Calculated effective mode volume, $V_{\text{eff}}$, at the center of the central bowtie as a function of bowtie width.}
    \label{fig:4}
\end{figure*}

\section*{Self-assembly of atomic-scale bowtie cavities}

 \begin{figure*}[t]
\centering
 \includegraphics[width=\textwidth]{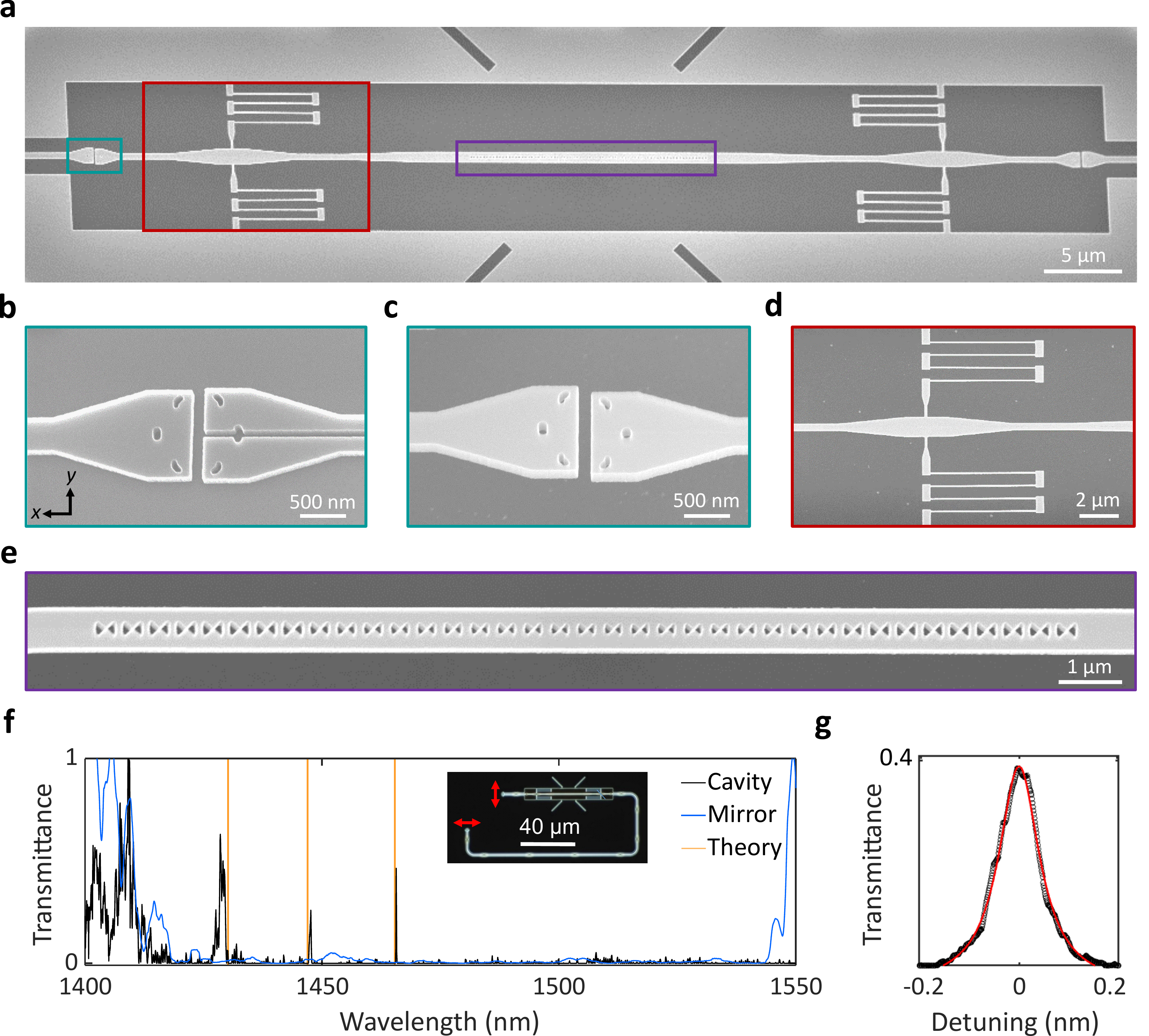}
    \caption{\textbf{Integration of self-assembled nanobeam cavities with photonic circuits.} \textbf{a}, Tilted-view ($20 ^{\circ}$) SEM image of a self-assembled nanobeam cavity terminated with photonic circuit crossings and including tapered waveguide sections and suspension springs. \textbf{b}, Tilted-view ($15 ^{\circ}$) SEM image of a circuit crossing before self-assembly.  \textbf{c}, Tilted-view ($25 ^{\circ}$) SEM image of a self-assembled circuit crossing. \textbf{d}, Top-view SEM image of the spring suspension attached to a tapered waveguide section. \textbf{e}, Tilted-view ($20 ^{\circ}$) SEM image of a self-assembled nanobeam cavity with 2 nm bowtie width. \textbf{f}, Transmission spectrum of a self-assembled nanobeam cavity normalized to the transmission of a reference structure based on a straight self-assembled waveguide. The transmission spectrum of a self-assembled bowtie nanobeam made entirely of mirror unit cells is shown for reference and to determine the mirror pass-bands. The inset shows a dark-field microscope image of the entire photonic circuit, which features orthogonally oriented (red arrows) broadband grating couplers. \textbf{g}, Lorentzian fit to the fundamental cavity resonance, yielding a quality factor of $1.5 \times 10^{4}$.}
    \label{fig:5}
\end{figure*}

\noindent
To illustrate the application of our method, we now turn to the realization of photonic nanocavities that confine light in air gaps in a 220 nm thick silicon membrane with aspect ratios exceeding 100. Figure~\ref{fig:2}a shows the geometry of a nanobeam cavity featuring a unit cell that includes a 2 nm air bowtie and is designed following well-known methods for nanobeam cavities (see Supplementary Section S2 for details on the cavity design). The normalized electric field of the fundamental optical mode is plotted respectively on a log-scale in Fig.~\ref{fig:2}b and on a linear scale, zooming-in on the central bowtie, in Fig.~\ref{fig:2}c. The fundamental cavity mode features a resonance wavelength of $\lambda = 1524$~nm, a quality factor of $Q = 5\times10^{4}$, and a mode volume of $V = 3.36\times10^{-4}\lambda^3$, calculated at the center of the central bowtie~\cite{lightningrod}. Note that the cavity design takes the constraints of our nanofabrication process into account(see Ref.~\cite{Marcus} and Supplementary Section S2), except for the use of a 2 nm air void at the bowtie centers, which is well below the resolution of current lithography technology and is therefore instead realized following the design rules provided by Fig. \ref{fig:1}d.

The nanobeam cavity is fabricated as two halves, each suspended by two folded cantilevers with a total spring constant of 0.038 N/m. The two halves are separated by a gap, $g_\text{f}$ = 50 nm, such that the set \{$k$,$A$,$g_\text{0}$\} lies deep within the parameter space leading to surface-force-assisted collapses (see Fig.~\ref{fig:1}d) and therefore the two halves adhere after underetching. Importantly, while the resolution of the nanofabrication limits the absolute value of $g_\text{f}$, the tip-to-tip distance before self-assembly, $g_\text{b}$, and thus the bowtie width in the final device, $g$, are limited only by surface roughness, enabling the realization of atomic-scale air bowties. We fabricate nanocavities with different bowtie widths by varying the offset, $\delta$, between $g_\text{f}$ and $g_\text{b}$ in the lithographic mask (see Supplementary Section S3.1 for details on the bowtie unit cell geometry). Figures~\ref{fig:2}d and e show two halves of a nanobeam cavity and their central bowtie unit cell before and after underetching, resulting in a bowtie width of approximately 2 nm for offset $\delta$ = 10 nm. Figure~\ref{fig:2}f shows a characteristic device, which includes 22 \micro m of unpatterned half-beams on each side of the photonic-crystal cavity to increase the surface forces and aid the self-assembly, as well as the two pairs of folded cantilever springs.

A systematic SEM study of devices with varying offset confirms the anticorrelation between offset and bowtie width (see Supplementary Section S3.2). However, few- or subnanometer gaps cannot be reliably measured with SEM, and we therefore turn to characterization using scanning transmission electron microscopy (STEM), (see Methods and Supplementary Section S3.3 for experimental details). Figure~\ref{fig:3}a and b show top-view annular dark-field STEM images of the central bowtie unit cell for self-assembled nanobeam cavities fabricated using $\delta$ = 10 nm and $\delta$ = 11 nm. We tilt the sample to align the electron beam to the [100] zone axis of the silicon membranes, and the (022) planes of the silicon crystal lattice with their characteristic inter-planar distance of 0.19 nm~\cite{lin2014phase_Sicrystalplanes} are visible, as shown in the high-resolution STEM images shown in Figs.~\ref{fig:3}c and d. By maximizing the intensity of the diffraction signal on both sides of the self-assembled bowtie, we find the (022) crystal planes in the top and bottom parts to be misaligned by 1-2$\degree$, e.g., 1.6$\degree$ and 1.8$\degree$ respectively in Figs.~\ref{fig:3}c and d, which is likely a consequence of minor deviations from perfect sidewall verticality or surface roughness. With these considerations in mind, we analyze the STEM images in Figs.~\ref{fig:3}c and d, acquired by keeping the bottom half of the bowtie normal to the incident electron beam. In Fig.~\ref{fig:3}c, the silicon (022) crystal plane of the upper and lower parts of the self-assembled bowtie are separated by a distance of 9.2 nm. The air bowtie is bounded on both sides by amorphous silicon oxide between the two crystalline regions, as confirmed by atomic composition analysis using electron energy-loss spectroscopy (EELS). The native silicon oxide has an estimated thickness, $d$, of 2 to 2.5 nm, which is in accordance with the native oxide on crystalline silicon devices~\cite{nasr2022effect}. Due to the small tilt angle of the top part of the nanobeam cavity, the STEM shows a smooth transition from the background through the oxide to silicon, while the lower part shows a sharper transition. The high-resolution STEM image of the central part of the bowtie for $\delta$ = 11 nm, shown in Fig.~\ref{fig:3}d, indicates that the two bowties are most likely touching at the native-oxide interface (see Supplementary Section S3.3 for STEM imaging on other offsets). This demonstrates the ability of our method to build atomic-scale semiconductor devices in which the critical dimension is limited by structural disorder rather than lithography. In addition, this opens up a whole new set of challenges in nanocavity research: Both the surface oxide and the surface roughness are generally considered to be irrelevant except for the impact on the quality factor, but here they play a decisive role in the mode volume and resonance frequency as well (see Supplementary Section S2.2). Interestingly, the low refractive index of silicon oxide relative to silicon enables glass-core bowtie nanocavities in which rare-earth ions can be implanted to form high-quality quantum emitters~\cite{Erbium_in_glass, galli2006strong}.

\section*{Optical characterization of self-assembled nanocavities and scalable integration with photonic circuits}
We characterize the resonant modes of the self-assembled nanobeam cavities by cross-polarized far-field resonant scattering, which results in Fano resonances due to the interference with a vertical mode of the structure~\cite{Marcus}. Representative spectra of sets of nominally identical cavities for offsets $\delta$ = 8, 9 and 10 nm are shown in Figs.~\ref{fig:4}a-c. The resonant wavelengths and quality factors of the cavity modes are extracted by fitting Fano lineshapes to the observed resonant features (see Supplementary Section S4.1). We measure quality factors between $7.8 \times 10^{3}$ and $3.9 \times 10^{4}$. The resonance wavelengths exhibit a clear red-shift with increasing offset, i.e., decreasing bowtie width. We estimate the average bowtie width in the measured cavities from the offset-to-width correspondence found via image analysis on a large set of SEM images of structures fabricated with $\delta$ = \{0,7,14,20\} nm (see Supplementary Section 3.2). The values for $\delta$ = 8, 9, and 10 nm correspond to $g$ = 5.3, 3.3 and 1.3 nm, respectively. We simulate the fabricated geometry, which includes a 2 nm native oxide layer (see Supplementary Section S2.2), for varying bowtie width and correlate the simulated and measured resonant wavelengths. Figure~\ref{fig:4}d shows both the measured and simulated resonant wavelength as a function of bowtie width, confirming the pronounced red shift with bowtie width. Spectral shifts between theory and experiment stemming from systematic errors in SEM measurements, local variations in the thickness of the device layer, etc.\ are commonly observed when analyzing nanocavities. In our case, we obtain an excellent agreement after red-shifting the theoretical curve by 15 nm. The measured $Q$-factors are smaller than the simulated values, as shown in Fig.~\ref{fig:4}e, mainly as a result of scattering losses due to structural disorder. Still, we consistently observe $Q$-factors exceeding previous experimental results on sub-diffraction confinement by more than an order of magnitude across multiple devices, even for 2 nm cavities that exhibit much smaller mode volumes than any previous experiments on dielectric cavities~\cite{Marcus,lightningrod} (see Fig.~\ref{fig:4}f). The high quality factors and the high process yield confirm the robustness of the surface-force self-assembly method.\\

Finally, we turn to the quest of interfacing self-assembled devices with complex circuitry, i.e., the scalability of our method for interfacing the bottom-up self-assembled devices with top-down planar technology. For photonic-crystal nanobeam cavities, the most well-known approaches are either evanescent side-coupling~\cite{sidecoupling} or (in-line) direct coupling~\cite{directcoupling}. These two strategies are less trivial to realize for self-assembled devices since they require efficient coupling between mechanically isolated self-assembled regions such as a nanobeam cavity and non-self-assembled regions such as suspended waveguides. To this end, we use a recently invented topology-optimized photonic component that enables a broadband waveguide-to-waveguide transmission window across a 100 nm air trench, which provides both electrical and mechanical isolation~\cite{Babak}. This enables the use of the self-assembly method by fabricating one of the sides of the component across the trench in two halves, which self-assembles at the same time as the nanobeam cavity.
A self-assembled nanobeam cavity, including efficient interfaces to external waveguides via such circuit crossings and low-loss anchor points for the springs on tapered waveguide regions is shown in Fig.~\ref{fig:5}a. Figures~\ref{fig:5}b and c show tilted SEM images of the circuit crossing before and after self-assembly. The in-plane directed self-assembly is accurate down to the resolution of the SEM. Still, some out-of-plane bowing is observed, which could readily be avoided by adding more springs or other means of stress-release management. As in the structures in Fig.~\ref{fig:2}, two sets of springs are used, but they are attached to the tapered waveguide section as shown in Fig.~\ref{fig:5}d. The taper works as a mode expander that minimizes the field intensity at the silicon edges, allowing a calculated transmission of 99.7 \% at the resonance wavelength of the cavity. Compared to the cavity shown in Fig.~\ref{fig:2}, the cavity for on-chip transmission experiments, which is shown in Fig.~\ref{fig:5}e, has a longer defect region to reduce out-of-plane radiation losses and a smaller number of mirror unit cells to facilitate efficient transmission through the cavity.

The photonic circuits beyond the crossings include two orthogonally oriented free-space grating couplers that allow measuring the circuit transmission through spatially resolved and cross-polarized spectroscopy~\cite{Christian_PTI_2022} as shown by the dark-field optical microscope image in the inset of Fig.~\ref{fig:5}f (see Supplementary Section S4.2 for an SEM image of a full device). The cavity transmittance is obtained by normalizing the measured transmitted power to that measured in a self-assembled suspended waveguide of equivalent length, i.e., all optical elements on the chip and in the optical setup are factored out (see Supplementary Section S4.2). Figure~\ref{fig:5}f shows the transmittance for two different self-assembled devices: First, a 2 nm air-bowtie nanobeam cavity with 8 mirror unit cells and, second, a 2 nm air-bowtie nanobeam mirror with 25 identical unit cells, both corresponding to structures with $\delta$ = 12 nm for the employed sample. The transmittance of the mirror is negligible between 1425 nm and 1540 nm, which is consistent with the simulated photonic band gap. The spectrum of the cavity device exhibits three distinct Lorentzian resonances which agree quantitatively with our numerical cavity model using the fabricated dimensions (2 nm air bowties and a 2 nm native oxide layer), provided the wavelength of the simulated eigenmodes are shifted by 11.6 nm.
This is well within the differences between experiments and theory commonly observed in nanophotonic devices based on the same top-down nanofabrication process~\cite{Christian_PTI_2022} and in close agreement with the 15 nm shift employed for Fig. \ref{fig:4}d. We observe a 39\% cavity transmittance, which is smaller than the simulated value of 96.4\% due to structural disorder (see Supplementary Section S4.2). The transmittance across the fundamental cavity mode is shown in Fig.~\ref{fig:5}g, and exhibits an irregular lineshape due to interference from reflection at the input and output grating couplers. By fitting to a Lorentzian lineshape, we obtain a $Q$ of $1.5$$\times$$10^4$, comparable to the values obtained in Fig.~\ref{fig:4} but with the notable difference that this is a loaded $Q$-factor and the cavity is efficiently coupled to a waveguide architecture. 

\section{Conclusion}

Our mapping of the phase space governing the collapse of suspended platforms provides a clear design rule both for new research aiming to exploit the deterministic self-assembly and for conventional micro- and nanoelectromechanical systems, where collapses are generally undesirable. Looking ahead, the introduction of a robust and accurate self-assembly method in planar technology opens perspectives for a wide range of research that seemed far beyond experimental reach until now. While we focus here on the role of atomic-scale nanometer void features, which may be used for solid-state nanopore sequencing~\cite{xue_solid-state_2020}, nanogap quantum tunneling electrodes for biosensing~\cite{he_sub-5_2022} or as ultra-high quality shadow-masks for superconducting quantum electronic devices~\cite{delaney_superconducting-qubit_2022}, lateral atomic-layer deposition before self-assembly might enable the formation of embedded atomic-scale structures for surface-enhanced Raman spectroscopy~\cite{luo2022high_nanogaps_selfassembly} or single-photon nonlinearities~\cite{raza2018ald_SERS_ALD_slotwaveguides,Choi}.
In this respect, our demonstration of optical cavities with atomic-scale features is the first step towards a new generation of nanophotonic and quantum photonic devices. For example, our self-assembled waveguide-coupled cavity features an unprecedented set of parameters: With a mode volume of $8.8$$\times$$10^{-4}$ cubic wavelengths and a loaded $Q$-factor of $1.5$$\times$$10^4$, the light-matter interaction is enhanced by a Purcell factor of $1.3$$\times$$10^6$ over a bandwidth of 14 GHz and with a high on-resonance transmission. By incorporating embedded emitters such as erbium-doped alumina deposited with atomic-layer deposition~\cite{Ronn_ALD_Er_doped_alumina_ACSphotonics_2016}, highly efficient single-photon sources at telecom wavelengths may be envisioned, possibly even with a high degree of quantum coherence due to the extreme Purcell enhancements. Such cavities may also enhance the bulk non-linearity of the embedded materials to a level where they could operate using single photons~\cite{Choi} and provide record single-photon optomechanical readout rates for gigahertz mechanical modes even in the absence of embedded materials~\cite{bozkurt_deep_2022}. More generally, our work opens perspectives for exploring new regimes of photonics, electronics, and mechanics at atomic scales while at the same time enabling scalable and self-aligned integration with large-scale chip architectures.

\quad

\section{Methods}
\quad
\subsection{Fabrication process}
The devices are fabricated on a commercial silicon-on-insulator substrate (Soitec) with a 220 nm-thick silicon device layer and a 2 \micro m-thick buried oxide layer. A two-layer hardmask is deposited on the silicon device layer, consisting of 30 nm poly-crystalline chromium and 12 nm poly-crystalline silicon layers, followed by a 50 nm layer of chemically semi-amplified resist (CSAR) applied by spin-coating. The patterns are exposed in the resist with a 100 keV 100 MHz JEOL9500FSZ electron-beam writer and transferred into the silicon device layer by a low-power switched reactive-ion etch. The buried oxide layer is selectively etched to suspend the devices with an anhydrous hydrofluoric-acid (99.995\%) vapour phase etcher (SPTS Primaxx uEtch), using ethanol as a catalyst. A process pressure of 131 Torr, and a slow etching recipe (etch rate of approximately 14 nm/min) are used for selective oxide etching. The fabrication process flow is detailed in Ref.~\cite{Marcus}, and the hardmask etching process is detailed in Refs.~\cite{Guillermo, Christian_PTI_2022}.

\subsection{Surface-force characterization}
The measurements in Fig.~\ref{fig:1} are performed on Sample A (1536 devices) with platforms of widths $w$ = [2, 3, 4, 5, 8, 16, 32, 64] \micro m with logarithmic variations of both $g_\text{f}$ and $k$ from 30 to 200 nm and from 0.0097 to 13 N/m, respectively. The silicon-on-insulator stack sets the thickness of the platform and height above the substrate to $h$ = 220 nm and $H$ = 2 \micro m, respectively. A 1 \micro m pitch, $w$ $\times$ $2$ array of 200 nm sidelength square is etched in the platform to facilitate the underetching. The devices also have trenches on the top-right and top-left of the platform to reduce potential fringing-field contributions to the surface forces~\cite{Fringingfields}. Scales are integrated on the right and left side of the platforms to measure displacements due to the built-in stress release, which imposes a baseline correction to the initial gap $g_\text{0}$ (see Supplementary Section S1.1). All these additional features have minimal effect on whether the devices collapse or not collapse.

\subsection{Scanning transmission-electron microscope}

Annular dark-field STEM imaging is performed using an FEI Titan 80-300 kV transmission electron microscope (TEM) operated at 300 kV to extract high-resolution images of the cavity bowtie region. The transmission electron microscope is fitted with a field-emission gun and an aberration-correction unit on the probe-forming lenses, giving it a spatial resolution better than 0.1 nm. A focused ion beam is used inside a FIB-SEM system (Helios Nanolab 600) to prepare the cavity structures for high-resolution imaging. A micromanipulator needle is welded to the cavity structure by induced deposition of Pt from a precursor source to transport the cavities from the sample to the TEM equipment. This is followed by cutting the tethers around the cavities using a Ga+ ion beam of 30 keV and 40 pA current, lifting the released cavities from the sample, relocating and welding the cavities to a TEM-compatible Cu grid, and finally detaching the micromanipulator needle from the cavities using the ion beam. 

\subsection{Optical measurements}

The optical spectrum of each nanocavity is measured using free-space confocal microscopy. Measurements are performed either by direct resonant scattering on isolated nanocavities (Fig.~\ref{fig:4}) or via transmission by coupling light in and out of photonic circuits with embedded nanocavities (Fig.~\ref{fig:5}). Two fiber-coupled tunable diode lasers (Santec TSL-710, $\lambda_1$ = 1355 - 1480 nm and $\lambda_2$ = 1480 - 1640 nm) are combined into a 4x1 optical switch (Santec OSU-110) for excitation of the nanocavities. Light is focused onto and collected from the sample using a 50X microscope objective (Mitutoyo Plan Apo NIR 50X, NA = 0.42). For resonant scattering measurements, the excitation and collection spots spatially overlap, and their polarizations are set at $45^{\circ}$ relative to the leading polarization of the cavity mode and orthogonal to each other (see Supplementary Section S4.1 for the schematic). For measurements of nanocavities embedded in photonic circuits, the excitation and collection are cross-polarized and spatially offset by employing two free-space grating couplers oriented orthogonal to each other. Both the grating couplers are kept 30 \micro m apart in vertical and horizontal directions. Spectra are acquired by sequentially sweeping the two tunable lasers (if needed) and detecting with a synchronous calibrated power meter (Santec MPM-210). The spectra are then normalized to the laser spectrum as measured with a direct patch fiber for resonant scattering measurements and to the spectrum of a suspended silicon waveguide of equivalent length for the photonic circuits.

\section{Acknowledgements}
We thank Marcus Albrechtsen for valuable discussions. The authors gratefully acknowledge financial support from the Villum Foundation Young Investigator Programme (Grant No. 13170),
Innovation Fund Denmark (Grant No. 0175-00022 -- NEXUS and Grant No. 2054-00008 -- SCALE),
the Danish National Research Foundation (Grant No. DNRF147 -- NanoPhoton),
Independent Research Fund Denmark (Grant No. 0135-00315 -- VAFL), the European Research Council (Grant. No. 101045396 -- SPOTLIGHT), and the European Union's Horizon 2021 research and innovation programme under a Marie Sklodowska-Curie Action (Grant No. 101067606 -- TOPEX).

\section{Author contributions}
A.N.B. and T.A.S. fabricated the devices and did the SEM characterization.
A.N.B. and G.A. did the optical characterization.
B.V.L, A.N.B, G.A. and K.T. did the numerical design and analysis.
S.K. did the STEM measurements.
A.N.B., T.A.S., and G.A. did the data analysis.
A.N.B., G.A., and S.S. prepared the manuscript with input from all authors.
S.S., B.V.L., and K.T. designed the experiment.
S.S. conceived, initiated and supervised the project with co-supervision by B.V.L and G.A.

\section{Data availability}
The data that supports this manuscript is available upon reasonable request.

\section{Competing financial interests}
The authors declare no competing financial interests.

\scriptsize

\bibliographystyle{naturemag-etalnoitalics.bst}
\bibliography{ref}

\end{document}


\title{Supplementary information for \\ Self-assembly of atomic-scale photonic cavities}

\author{Ali Nawaz Babar}
\email{anaba@dtu.dk}
\affiliation{Department of Electrical and Photonics Engineering, DTU Electro, Technical University of Denmark, Building 343, DK-2800 Kgs. Lyngby, Denmark}
\affiliation{NanoPhoton - Center for Nanophotonics, Technical University of Denmark, {\O}rsteds Plads 345A, DK-2800 Kgs. Lyngby, Denmark.}

\author{Thor August Schimmell Weis}
\affiliation{Department of Electrical and Photonics Engineering, DTU Electro, Technical University of Denmark, Building 343, DK-2800 Kgs. Lyngby, Denmark}

\author{Konstantinos Tsoukalas}
\affiliation{Department of Electrical and Photonics Engineering, DTU Electro, Technical University of Denmark, Building 343, DK-2800 Kgs. Lyngby, Denmark}

\author{Shima Kadkhodazadeh }
\affiliation{NanoPhoton - Center for Nanophotonics, Technical University of Denmark, {\O}rsteds Plads 345A, DK-2800 Kgs. Lyngby, Denmark.}
\affiliation{DTU Nanolab, Technical University of Denmark, Building 347,
DK-2800 Kgs. Lyngby, Denmark}

\author{Guillermo Arregui}
\affiliation{Department of Electrical and Photonics Engineering, DTU Electro, Technical University of Denmark, Building 343, DK-2800 Kgs. Lyngby, Denmark}

\author{Babak Vosoughi Lahijani}
\affiliation{Department of Electrical and Photonics Engineering, DTU Electro, Technical University of Denmark, Building 343, DK-2800 Kgs. Lyngby, Denmark}
\affiliation{NanoPhoton - Center for Nanophotonics, Technical University of Denmark, {\O}rsteds Plads 345A, DK-2800 Kgs. Lyngby, Denmark.}

\author{S{\o}ren Stobbe}
\email{ssto@dtu.dk}
\affiliation{Department of Electrical and Photonics Engineering, DTU Electro, Technical University of Denmark, Building 343, DK-2800 Kgs. Lyngby, Denmark}
\affiliation{NanoPhoton - Center for Nanophotonics, Technical University of Denmark, {\O}rsteds Plads 345A, DK-2800 Kgs. Lyngby, Denmark.}

\maketitle

\section{S1. Investigation of pull-in instabilities due to surface forces} \label{sec:surfaceforces}

\begin{figure}[t]
 \centering
\includegraphics[width=0.88
\columnwidth]{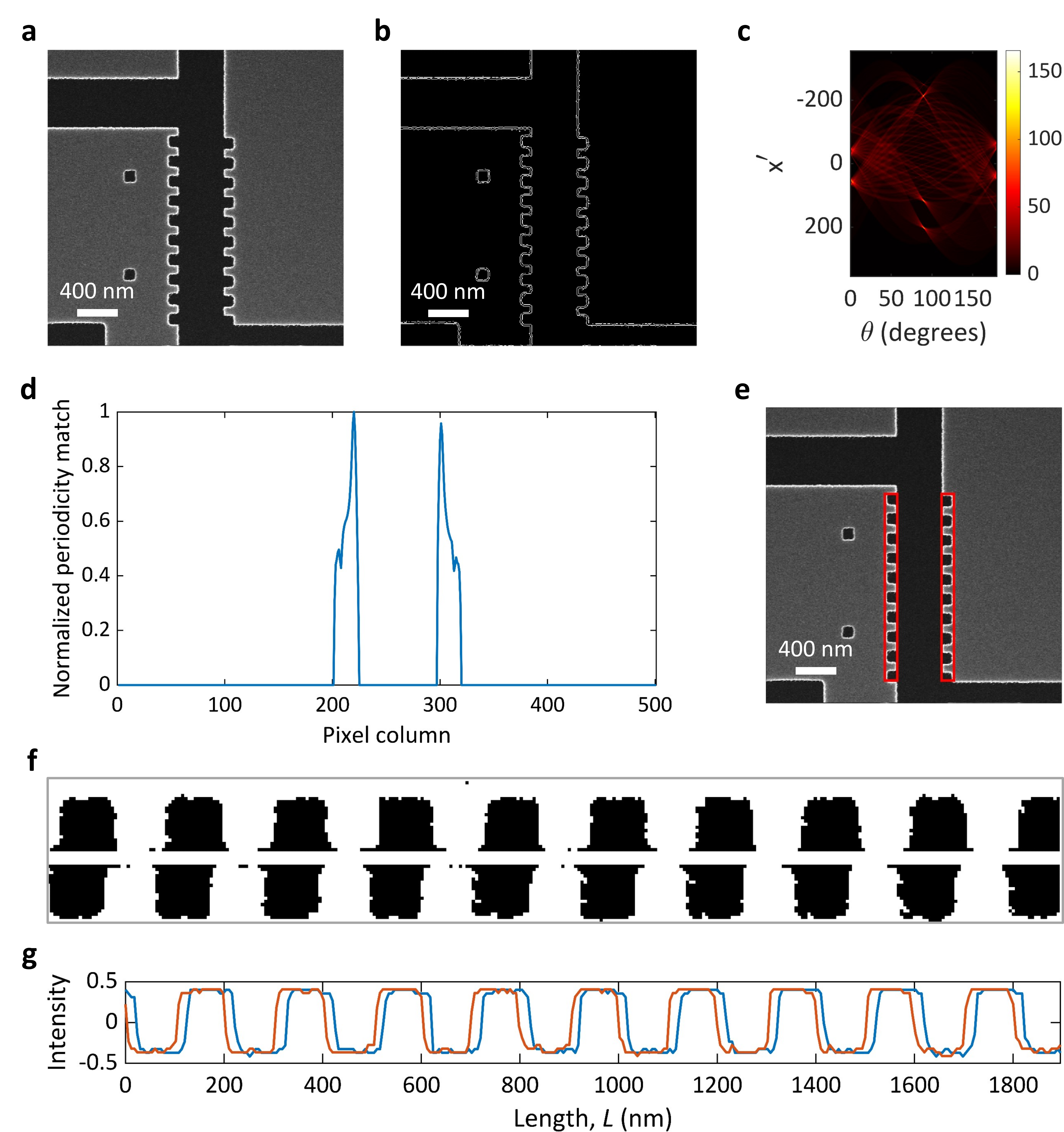}
\caption[]{\textbf{Automatic image-analysis of relative scale displacement.} \textbf{a}, Top-view SEM image of a scale attached to the platform with a fabricated gap, $g_\text{f}$, of 520 nm (after underetching). \textbf{b}, Sobel edge detection. \textbf{c}, Radon transform of the image to measure the device's rotation angle relative to the image's coordinate system. $\Theta$ is the angle of rotation of an axis centered on the image. $x'$ is the distance in pixels of a given pixel from the rotated axis. \textbf{d}, Match between periodicity found in the image columns and the known periodicity of the scales. \textbf{e}, Detected scales. \textbf{f}, Scales cropped and converted to a black-and-white image. \textbf{g}, Graph obtained by summing the pixel counts of the black-and-white image, normalizing, and centering.}
\label{fig:stressrelease}
\end{figure}

\subsection{S1.1. Displacement measurements}
\noindent The built-in stress in the device layer of silicon-on-insulator wafers causes expansion or contraction when the silicon oxide is selectively etched away to release the structures. This means that the initial gaps in the platforms we investigate, $g_\text{0}$, i.e., the gaps as they would be without the surface forces, do not exactly correspond to the fabricated gaps before underetching, $g_\text{f}$. The initial gap is modified by the stress-release displacement, $\Delta g$, which we correct for in our experiments as explained below. This displacement is experimentally obtained by including, for each spring constant and platform width, a reference platform with a fabricated 520 nm gap, which is large enough to diminish the surface forces by orders of magnitude~\cite{Gusso_DispersionForces}. We acquire scanning electron microscope (SEM) images of all such devices and perform image analysis to extract $\Delta g$. As mentioned in the Methods section of the main text, the platforms are equipped with scales on the sides, see Fig.~\ref{fig:stressrelease}a, allowing us to accurately extract their displacement from an SEM image acquired at high scan speed. The scales are analyzed by first using Sobel edge detection as shown in Fig.~\ref{fig:stressrelease}b, followed by a Radon transform as shown in Fig.~\ref{fig:stressrelease}c. The latter detects the angle of rotation of the device relative to the coordinate system of the image and compensates for it by counter-rotating with a bi-linear rotation. The pixel columns containing the scales are found by column-wise fast Fourier transforms (FFT) and identification of the Fourier components matching the known periodicity of the scales as shown in Fig.~\ref{fig:stressrelease}d. The extracted pixel columns are then limited to the rows containing periodic structures, leaving only the areas marked in Fig.~\ref{fig:stressrelease}e. The detected scales are cropped from the image and converted to black and white images as seen in Fig.~\ref{fig:stressrelease}f, from which the number of black pixels in each column is counted to give the normalized plot seen in Fig.~\ref{fig:stressrelease}g. The phase difference between the two resulting curves is found by Fourier analysis. Based on the measurements of 57 devices, we find a mean stress-release displacement of 19.4 nm with a standard deviation of 2.8 nm, independent of the platform width and the spring constant. Finally, the initial gap, $g_\text{0}$, used in Fig.~1d of the main text, is obtained by systematically acquiring SEM images of the platforms before removing the buried oxide layer to extract the fabricated gap, $g_\text{f}$, followed by subtraction of the known displacement due to stress release, $\Delta g$, such that the initial gap is extracted as $g_\text{0} = g_\text{f}-\Delta g$.
\label{subsec:stressrelease}

\begin{figure}[ht]
 \centering
\includegraphics[width=0.9\columnwidth]{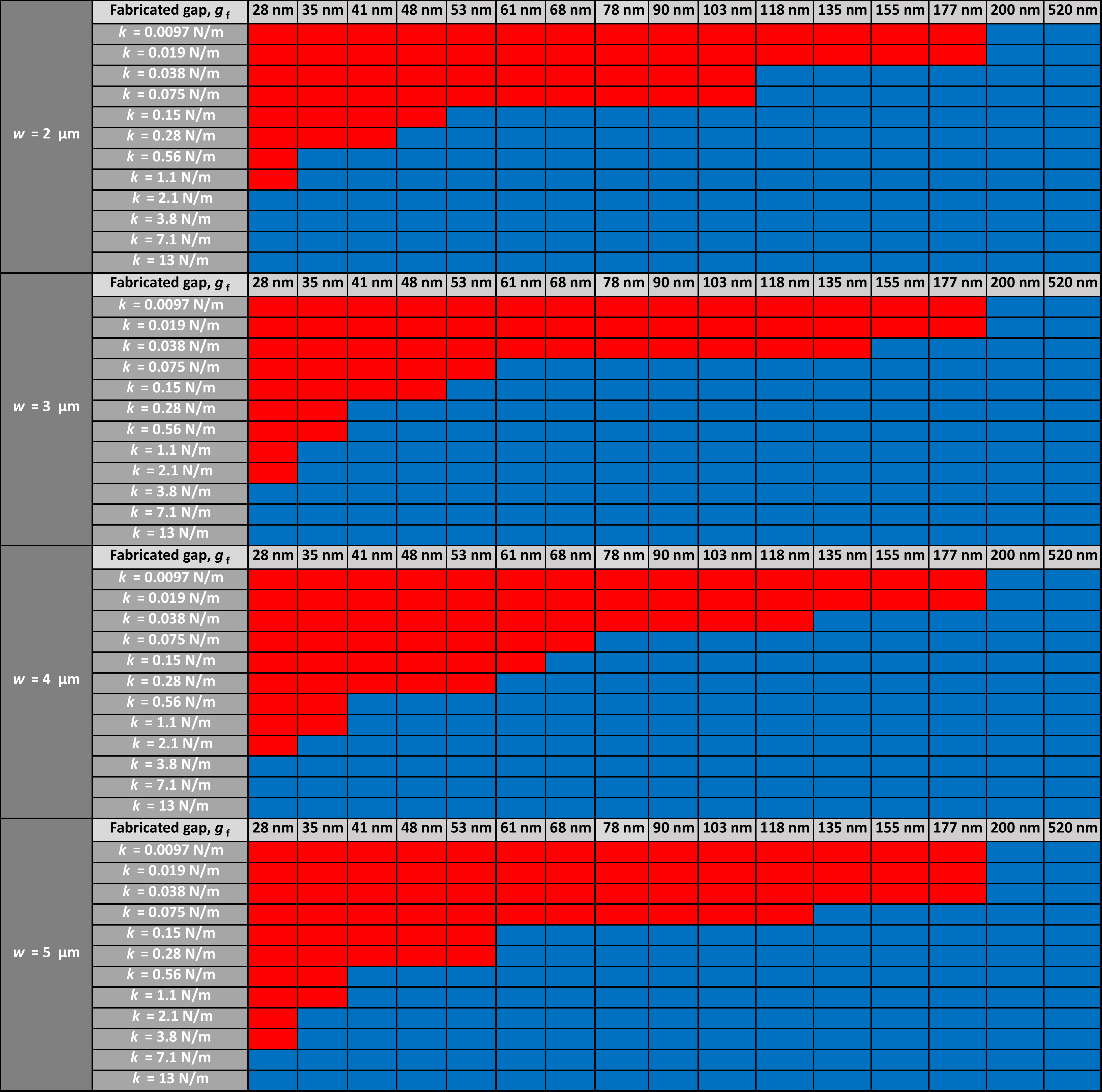}
\caption[]{\textbf{Post-underetching structural state of silicon platforms with widths 2, 3, 4, and 5 \micro m, respectively, for Sample A.} The columns indicate platforms with a fabricated gap, $g_\text{f}$, measured before releasing the structures, and the rows indicate platforms with different spring constants, $k$. Each block shows the experimental data for a specific platform width, $w$. The red cells represent platforms that collapsed in-plane on the anchored silicon, and the blue cells indicate platforms that did not collapse.}
\label{fig:Raw_data_1}
\end{figure}

\begin{figure}[h!]
 \centering
\includegraphics[width=0.9\columnwidth]{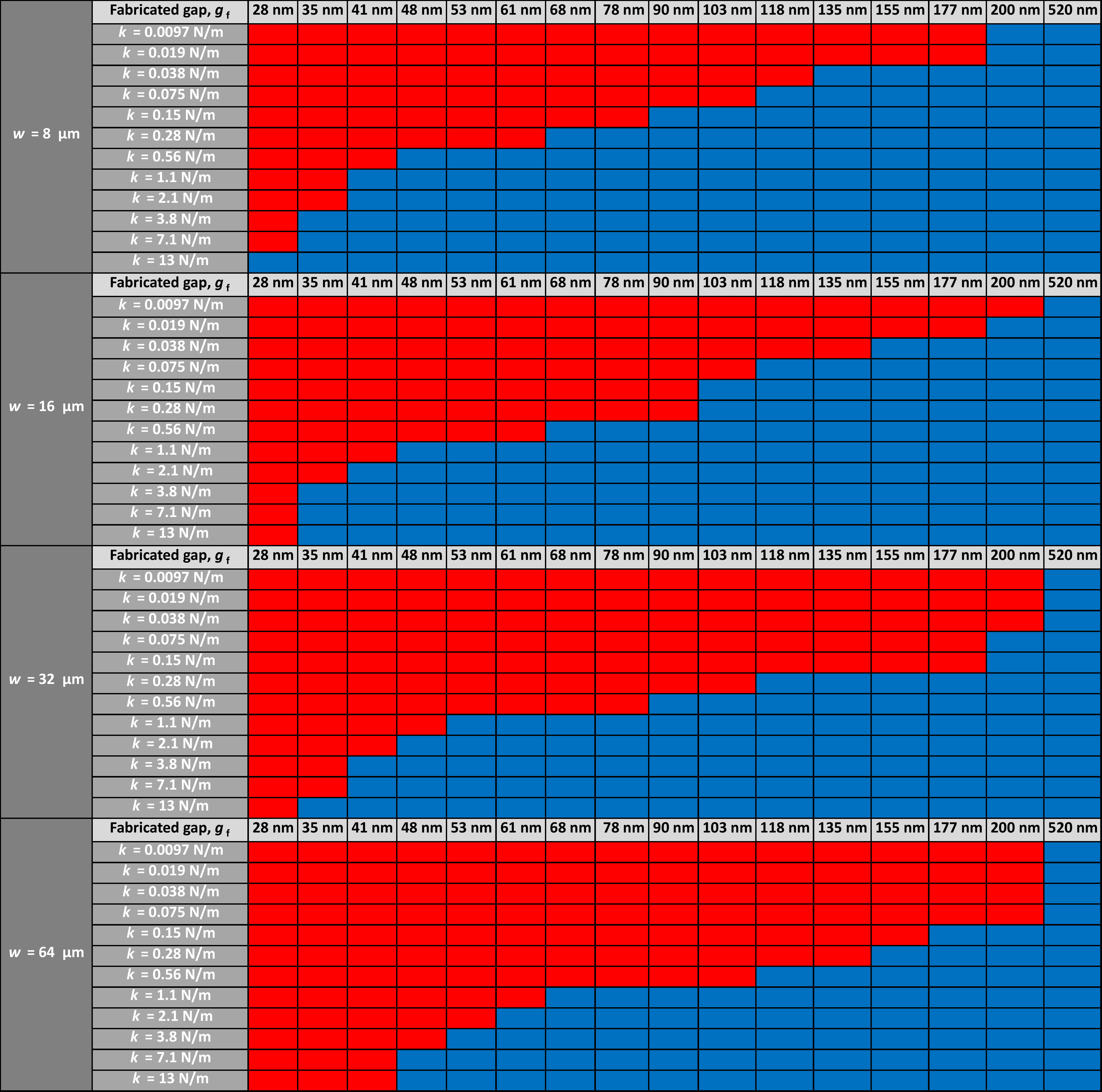}
\caption[]{\textbf{Post-underetching structural state of silicon platforms with widths 8, 16, 32, and 64 \micro m, respectively, for Sample A.} The columns indicate platforms with a fabricated gap, $g_\text{f}$, measured before releasing the structures, and the rows indicate platforms with different spring constants, $k$. Each block shows the experimental data for a specific platform width, $w$. The red cells represent platforms that collapsed in-plane on the anchored silicon, and the blue cells indicate platforms that did not collapse.}
\label{fig:Raw_data_2}
\end{figure}

\begin{figure}[ht]
 \centering
\includegraphics[width=0.9\columnwidth]{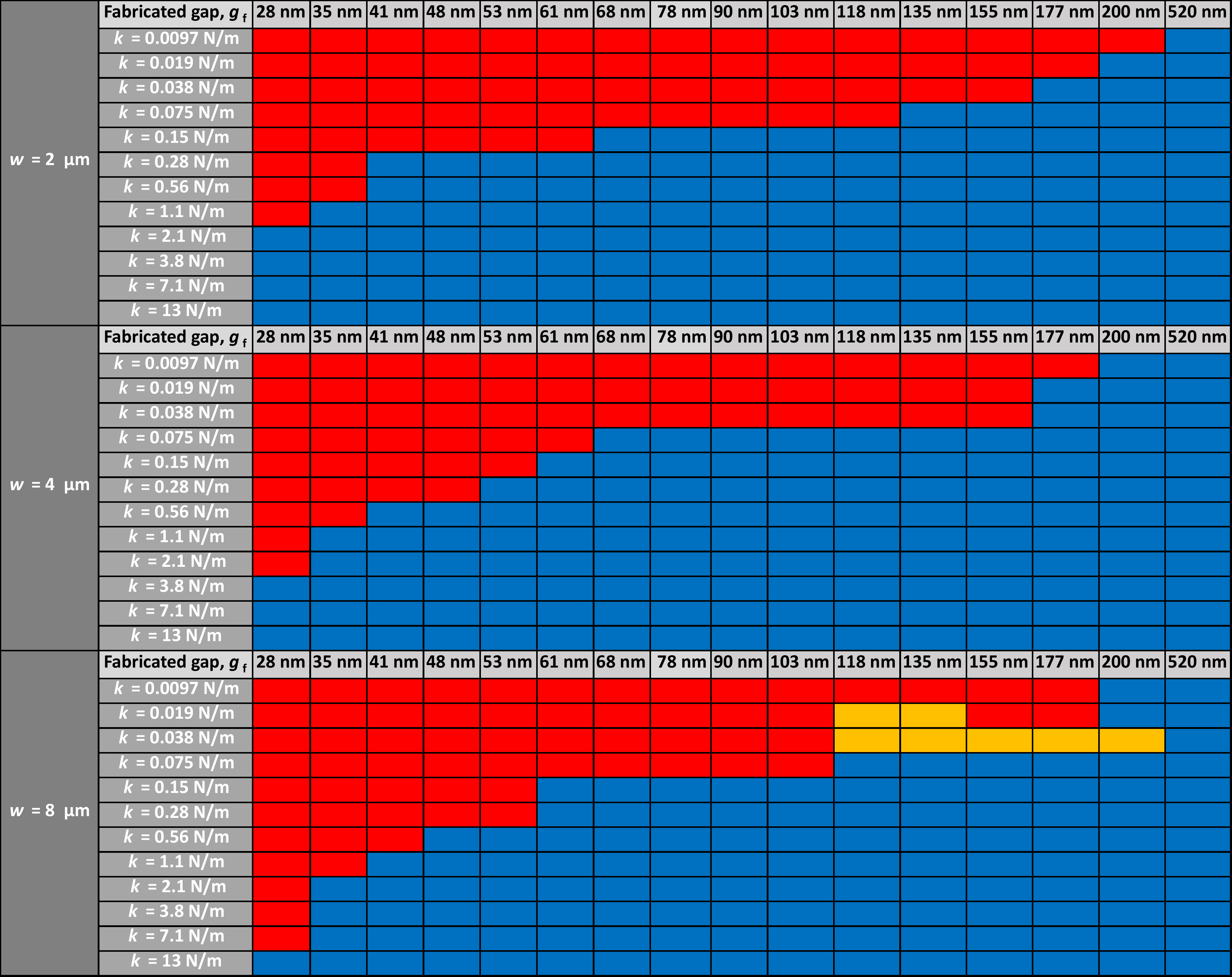}
\caption[]{\textbf{Post-underetching structural state of silicon platforms with widths 2, 4, and 8 \micro m, respectively, for Sample B.} The columns indicate platforms with a fabricated gap, $g_\text{f}$, measured before releasing the structures, and the rows indicate platforms with different spring constants, $k$. Each block shows the experimental data for a specific platform width, $w$. The red cells represent platforms that collapsed in-plane on the anchored silicon, and the blue cells indicate platforms that did not collapse. The yellow cells indicate the devices excluded due to fabrication imperfections or spring failure.}
\label{fig:Raw_data_3}
\end{figure}

\begin{figure}[ht]
 \centering
\includegraphics[width=0.9\columnwidth]{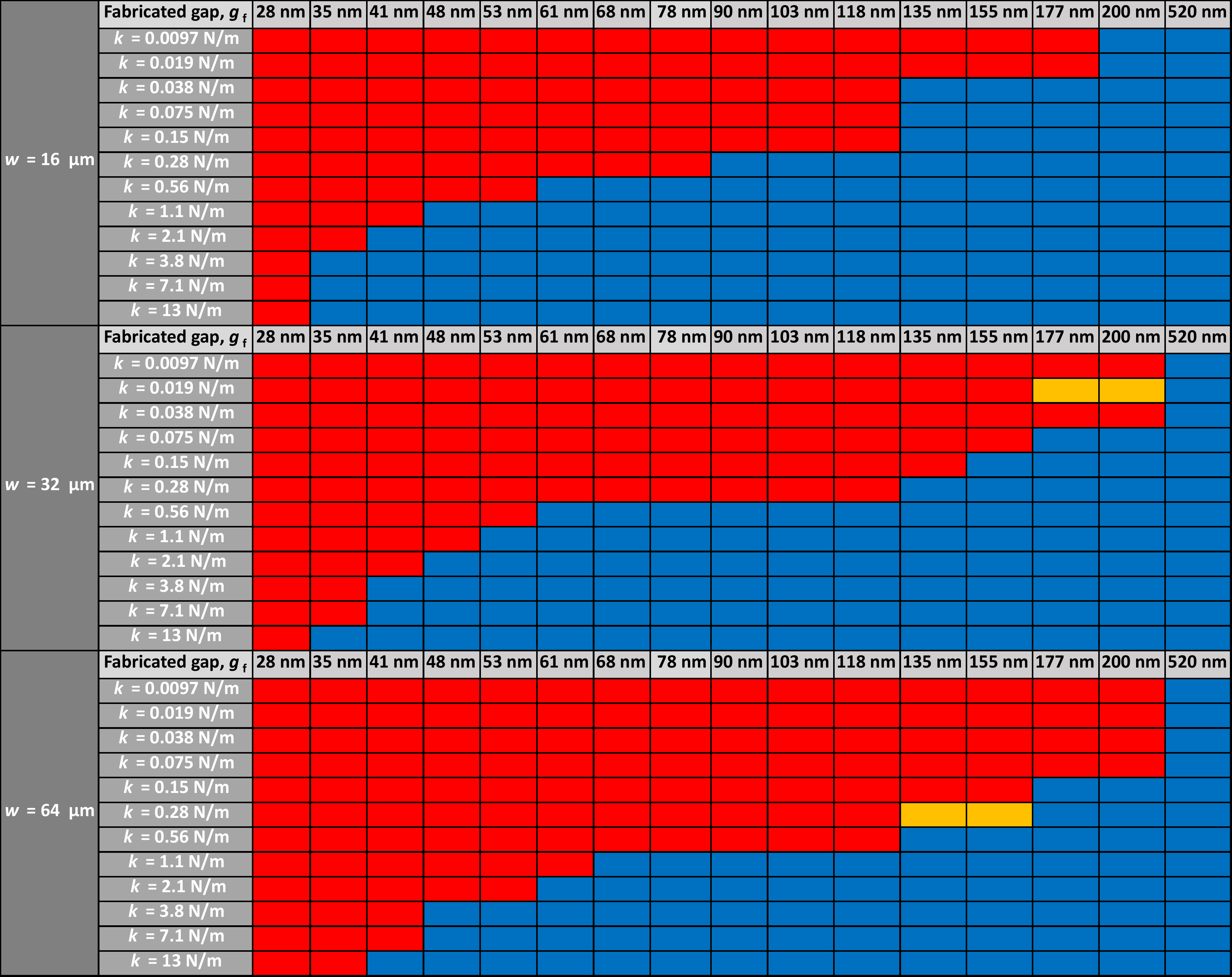}
\caption[]{\textbf{Post-underetching structural state of silicon platforms with widths 16, 32, and 64 \micro m, respectively, for Sample B.} The columns indicate platforms with a fabricated gap, $g_\text{f}$, measured before releasing the structures, and the rows indicate platforms with different spring constants, $k$. Each block shows the experimental data for a specific platform width, $w$. The red cells represent platforms that collapsed in-plane on the anchored silicon, and the blue cells indicate platforms that did not collapse. The yellow cells indicate the devices excluded due to fabrication imperfections or spring failure.}
\label{fig:Raw_data_4}
\end{figure}

\begin{figure}[ht]
 \centering
\includegraphics[width=0.8\columnwidth]{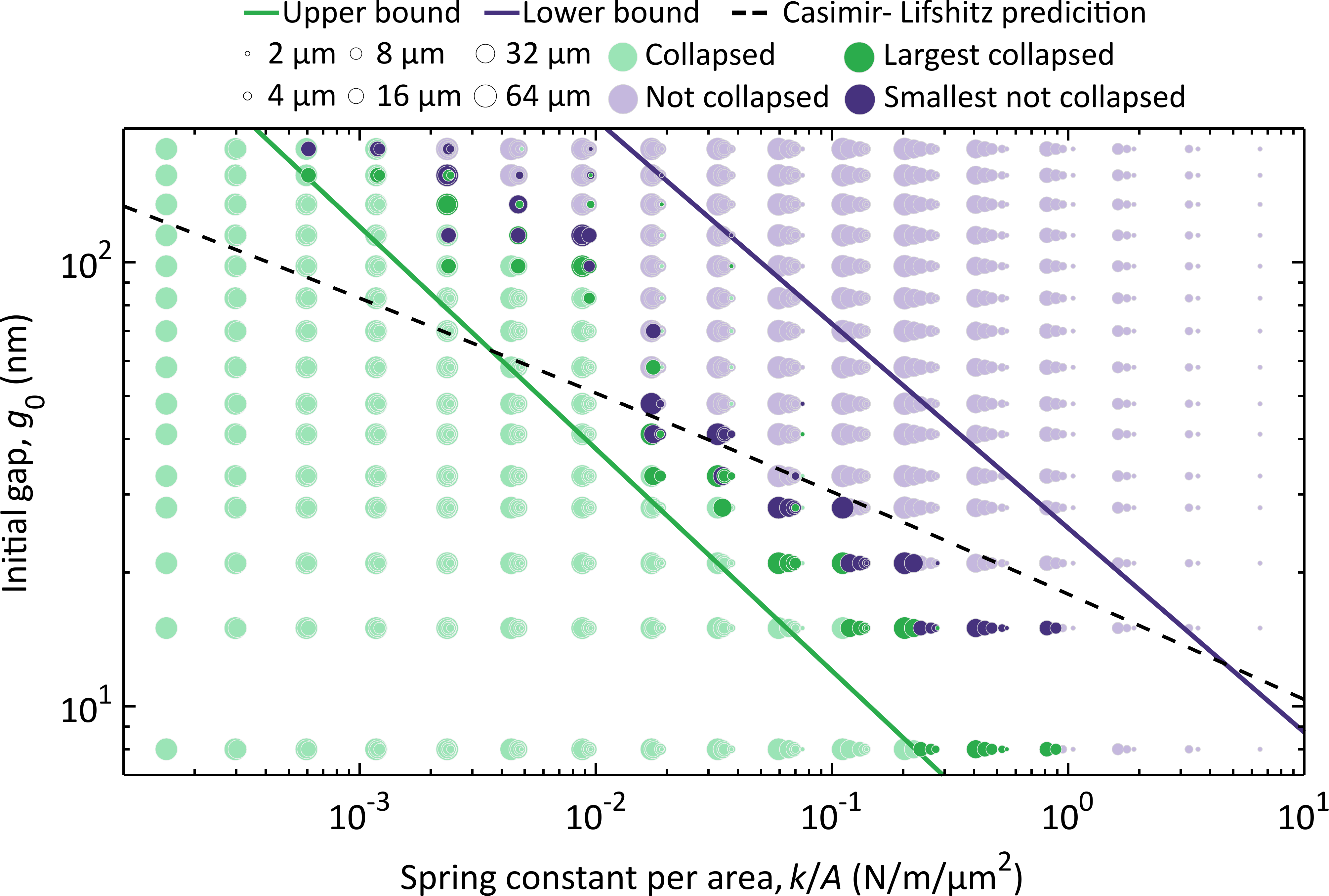}
\caption[]{\textbf{Measured map of the design space for self-assembly with compliant silicon structures.} The map is obtained by characterizing 1152 platforms by SEM (Sample B). The different sizes of the circles represent the
different widths of the platforms. The dark purple circle indicate the smallest initial gap not leading to a collapsed platform and the dark green circle indicates the largest initial gap leading to a collapsed platform. All the devices below the upper bound collapse, while those above the lower bound do not.}
\label{fig:SampleB}
\end{figure}

\subsection{S1.2. Mapping the threshold for self-assembly by surface forces: Raw data}

We fabricate a total of 2688 devices distributed across two samples (Sample A: 1536 devices; Sample B: 1152 devices) with different values of platform width, $w$, fabricated gap, $g_\text{f}$, and spring constant, $k$, as discussed in the Methods section. We characterize the devices using SEM performed with a single high-speed scan to minimize charge-induced displacement of the platforms, except when imaging for illustrative purposes, as in Fig.~1c in the main text.  Figures~\ref{fig:Raw_data_1} and \ref{fig:Raw_data_2} show the resulting map of the devices that either collapsed in-plane (red) on the anchored silicon or did not collapse (blue). The devices that failed due to fabrication imperfections or spring failure, such as out-of-plane collapse (yellow), are excluded from the final dataset. The data shown in Figs.~\ref{fig:Raw_data_1} and \ref{fig:Raw_data_2} is used to plot Fig.~1d in the main text. This experiment is repeated on Sample B to verify the reproducibility and robustness of our approach, and the raw data is shown in Figs.~\ref{fig:Raw_data_3} and \ref{fig:Raw_data_4}. Figure.~\ref{fig:SampleB} shows the self-assembly design space with compliant silicon structures for Sample B (1152 devices) . In this case, we find that all platforms for which $g_0 < 4.2(k/A)^{-0.48}$ collapse and all platforms for which $g_0 > 24(k/A)^{-0.47}$ do not collapse.

\subsection{S1.3. The critical gap of Casimir-Lifshitz theory as the threshold for self-assembly by surface forces}

The van der Waals force, which is responsible for surfaces adhering when they touch, is the short-distance (non-retarded) limit of the more general Casimir-Lifshitz force that in the idealized case of perfectly reflecting infinitely extended surfaces reduces to the long-range attractive force described by Casimir~\cite{casimir1948attraction}. When the intervening material between two surfaces is vacuum, the Casimir-Lifshitz force is attractive and increases non-linearly with decreasing separation, $g$, between the surfaces. If two surfaces are initially separated by the critical gap, $g_\text{c}$, the Casimir-Lifshitz force can overwhelm the elastic forces that hold the surfaces in place, bring them in contact, and subsequently stick them onto each other. Here, we use a lumped-element model where the elastic force is described by a linear spring with spring constant $k$ to calculate the critical gap.

The total force between two parallel surfaces separated by a gap $g=g_0-x$, where $g_0$ is the initial gap between the surfaces and $x$ is their displacement due to the Casimir-Lifshitz attraction is given by
\begin{equation}
    \label{eq:Ftot}
    F_\text{tot}(g)= A F_\text{CL}(g) - k(g_0-g),
\end{equation}
where $A$ is the area and $F_\text{CL}(g)$ is the Casimir-Lifshitz force per unit area between two surfaces given by \cite{Gusso_DispersionForces} 
\begin{equation}
    \label{eq:CLintegral}
    \begin{split}
    F_\text{CL}(g) & = -\frac{\hbar}{2 \pi^2 c^3}  \int_1^\infty p^2 \text{d}p 
   \int_0^\infty  \xi^3 \text{d} \xi \Bigg\{    \left[ \left( \frac{K+\epsilon \left( i \xi \right)p   }{K-\epsilon \left( i \xi \right)p } \right)^2 e^{2(\xi/c)pg}-1 \right]^{-1} \\
  & + \left[ \left( \frac{K+p}{K-p } \right)^2 e^{2(\xi/c)pg}-1 \right]^{-1}
  \Bigg\},
    \end{split}
\end{equation}
with $K = \sqrt{p^2 -1 + \epsilon \left( i \xi \right)}$ and the dielectric function of silicon at imaginary frequencies given by $\epsilon \left( i \xi \right)= 1+ \frac{10.703}{1+ \left(  \xi*1.506*10^{-16}  \right)^{1.83}}$, where $\xi$ is measured in rad/s~\cite{moazzami2021self}. The system loses linear stability at the point $g^*$ when
\begin{equation}
    \label{eq:stab1}
    \frac{\partial  F_\text{tot}(g^*)}{ \partial g} = 0.
\end{equation}
The critical gap, $g_c$, is defined as the initial gap that leads to the instability point, $g^*$, after the surfaces have been attracted by the Casimir force, i.e.,
\begin{equation}
    \label{eq:stab2}
    g_c=\frac{A}{k}F_\text{CL}(g^*)+g^*.
\end{equation}
To calculate the critical gap $g_c$ for a given $k/A$, we first numerically solve Eq.~(\ref{eq:stab1}) for $g^*$ which we subsequently insert in Eq.~(\ref{eq:stab2}).

\section{S2. Design of nanobeam photonic-crystal cavities with bowtie unit cells}
\label{sec:cavitydesign}

\begin{figure}[t]
\centering
\includegraphics[width=1\textwidth]{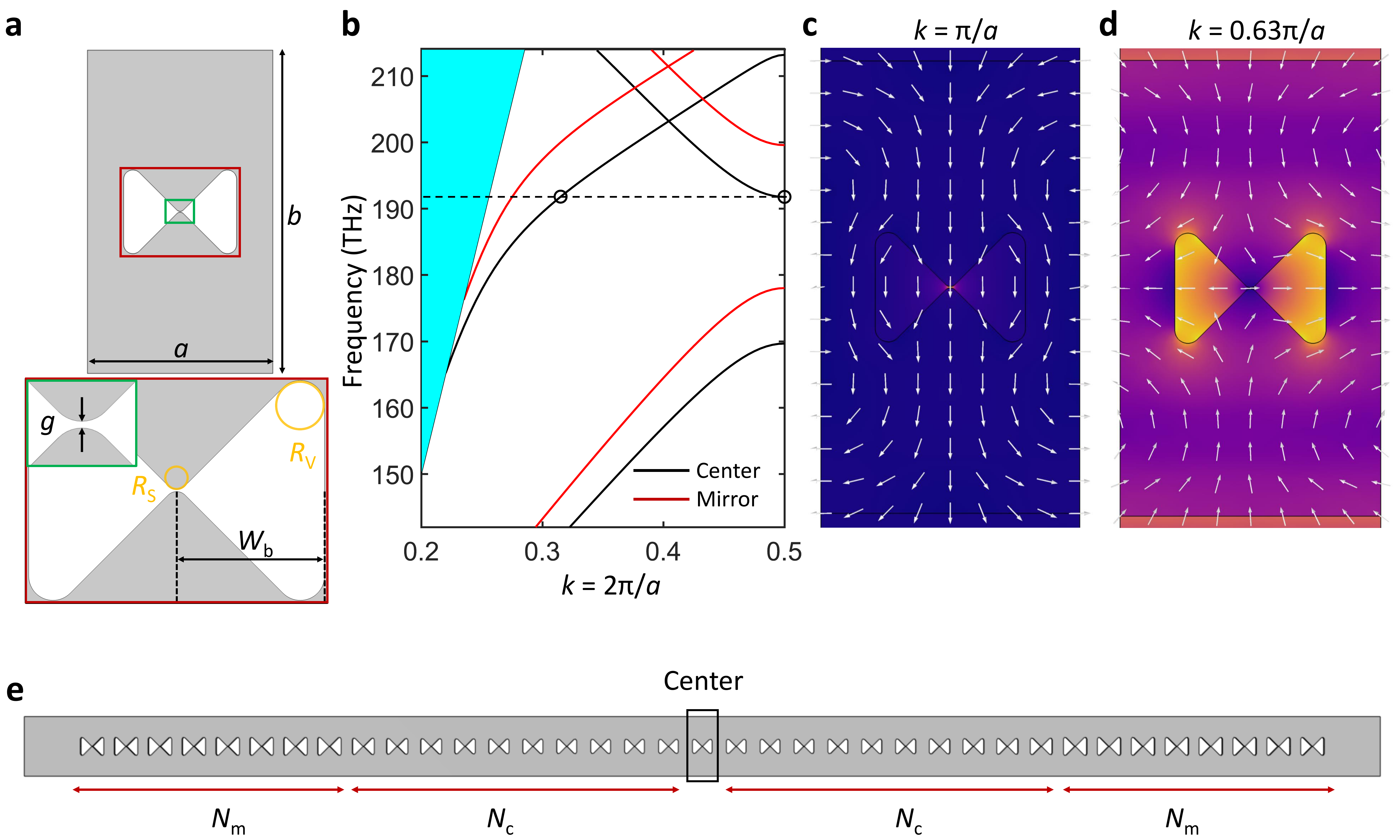}
\caption[]{\textbf{Design of a photonic-crystal nanobeam bowtie cavity.}  \textbf{a}, Geometric parameters of a single bowtie unit cell with an air bowtie of width $g$. \textbf{b}, Band structure (TE-like modes) for the central unit cell ($W_\text{b}$ = 116 nm, black curve), and for the mirror unit cell ($W_\text{b}$ = 139 nm, red curve). The blue region indicates the light cone. \textbf{c,d}, normalized $|\mathbf{E}|$-field of the center bowtie unit cell in the middle plane of the structure and at the edge of the Brillouin zone (\textbf{c}), and at a different wavenumber but the same frequency (\textbf{d}). White arrows indicate the electric field direction at the center of the silicon slab. \textbf{e}, Schematic of the photonic-crystal nanobeam bowtie cavity comprised of a central unit cell, tapering unit cells, $N_\text{c}$, and mirror unit cells, $N_\text{m}$.}
\label{fig:Cavity_design}
\end{figure}

\subsection{S2.1 Band structures and cavity design}
\label{subsec:bands}

The nanobeam photonic-crystal cavities we explore for self-assembly have a unit cell with a single-digit nanometer bowtie width at the center. Figure~\ref{fig:Cavity_design}a shows a triangular bowtie unit cell with a nanobeam width of $b$ = 700 nm, a lattice constant of $a$ = 400 nm, an air-bowtie width of $g$ = 2 nm, and a triangle width of $W_\text{b}$ = 116 nm. The bowtie angle is fixed to 90 degrees. The measured fabrication constraints, i.e., the smallest solid ($R_\text{S}$ = 10 nm) and void ($R_\text{V}$ = 20 nm) radii of curvature, are included in the cavity design to comply with our fabrication process, as shown in Fig.~\ref{fig:Cavity_design}a  . The band diagram of the unit cell is computed using a finite-element method in COMSOL Multiphysics and shown with solid black lines in Fig.~\ref{fig:Cavity_design}b, indicating that the second-order mode at the Brillouin-zone edge is a bowtie mode (Fig.~\ref{fig:Cavity_design}c). Fig.~\ref{fig:Cavity_design}b also includes the band structure for a unit cell with slightly modified parameters ($b$ = 700 nm, $a$ = 400 nm, $g$ = 2 nm, $W_\text{b}$ = 139 nm), which we define to be the mirror unit cell of our nanobeam cavity. We observe that the bowtie band at the edge of the Brillouin zone resides within the band gap of the mirror unit cell, which allows the generation of spatially-confined cavity modes by adiabatic tapering of the mirror unit cell into the center unit cell and back into the mirror unit cell. The band diagrams show that another mode (Fig.~\ref{fig:Cavity_design}d) is present at the same energy but at a lower wavenumber. These two modes, as shown with the superimposed vector fields in Figs.~\ref{fig:Cavity_design}c and d, have different symmetries, which ensures that a cavity design respecting the symmetry cannot couple the bowtie mode to the other modes. The exact cavity geometry is built following a well-known procedure~\cite{MarkoLoncar} in which the unit cell is continuously transformed along $N_\text{c}$ unit cells in both directions from the center to the mirror unit cell, followed by $N_\text{m}$ mirror unit-cell sections. Figure \ref{fig:Cavity_design}e shows the cavity geometry used for the waveguide-coupled photonic-crystal nanobeam cavities, which uses $N_\text{c}$ = 10 and $N_\text{m}$ = 8. Table.~\ref{tab:Waveguidecoupledcavity} shows the geometric parameters for the waveguide-coupled photonic crystal nanobeam cavity, notably the value of $W_\text{b}$ at the $i$-th defect unit cell, $W_\text{b, i}$. On the contrary, the cavity geometry used for the resonant scattering measurements of Fig. 4 in the main text uses a much shorter defect region with $N_\text{c}$ = 3 and $N_\text{m}$ = 16 and Table.~\ref{tab:Cavity_Scattering} shows the geometric parameters for that cavity.

\begin{table}[h]
\begin{tabular}{|c|c|}
\hline
\textbf{Parameter}  & \textbf{\quad Dimensions (nm) \quad} \\ \hline
Nanobeam width, $b$   & 700             \\ \hline
\quad Lattice constant, $a$ \quad & 400             \\ \hline
Air-bowtie width, $g$ & 2               \\ \hline
$W_\text{b,1}$           & 116             \\ \hline
$W_\text{b,2}$           & 116             \\ \hline
$W_\text{b,3}$           & 117             \\ \hline
$W_\text{b,4}$           & 117             \\ \hline
$W_\text{b,5}$          & 118             \\ \hline
$W_\text{b,6}$           & 119             \\ \hline
$W_\text{b,7}$          & 120             \\ \hline
$W_\text{b,8}$          & 122             \\ \hline
$W_\text{b,9}$          & 124             \\ \hline
$W_\text{b,10}$         & 126             \\ \hline
$W_\text{b,11}$         & 130             \\ \hline
$W_\text{b,12}$  - $W_\text{b,19}$         & 139             \\ \hline
\end{tabular}
\caption{Geometric parameters for the waveguide-coupled photonic crystal nanobeam cavity}
\label{tab:Waveguidecoupledcavity}
\end{table}

\begin{table}[]
\begin{tabular}{|c|c|}
\hline
\textbf{Parameter}     & \textbf{\quad Dimensions (nm) \quad} \\ \hline
\quad Nanobeam width, $b$  \quad    & 700                      \\ \hline
Lattice constant, $a$    & 400                      \\ \hline
Air-bowtie width, $g$    & 2                        \\ \hline
$W_\text{b,1}$             & 119                      \\ \hline
$W_\text{b,2}$            & 120                      \\ \hline
$W_\text{b,3}$               & 122                      \\ \hline
$W_\text{b,4}$               & 127                      \\ \hline
$W_\text{b,5}$  - $W_\text{b20}$  & 141                      \\ \hline
\end{tabular}
\caption{Geometric parameters for the photonic crystal nanobeam cavity for the resonant scattering measurements}
\label{tab:Cavity_Scattering}
\end{table}

\begin{figure}[t]
\centering
\includegraphics[width=1\textwidth]{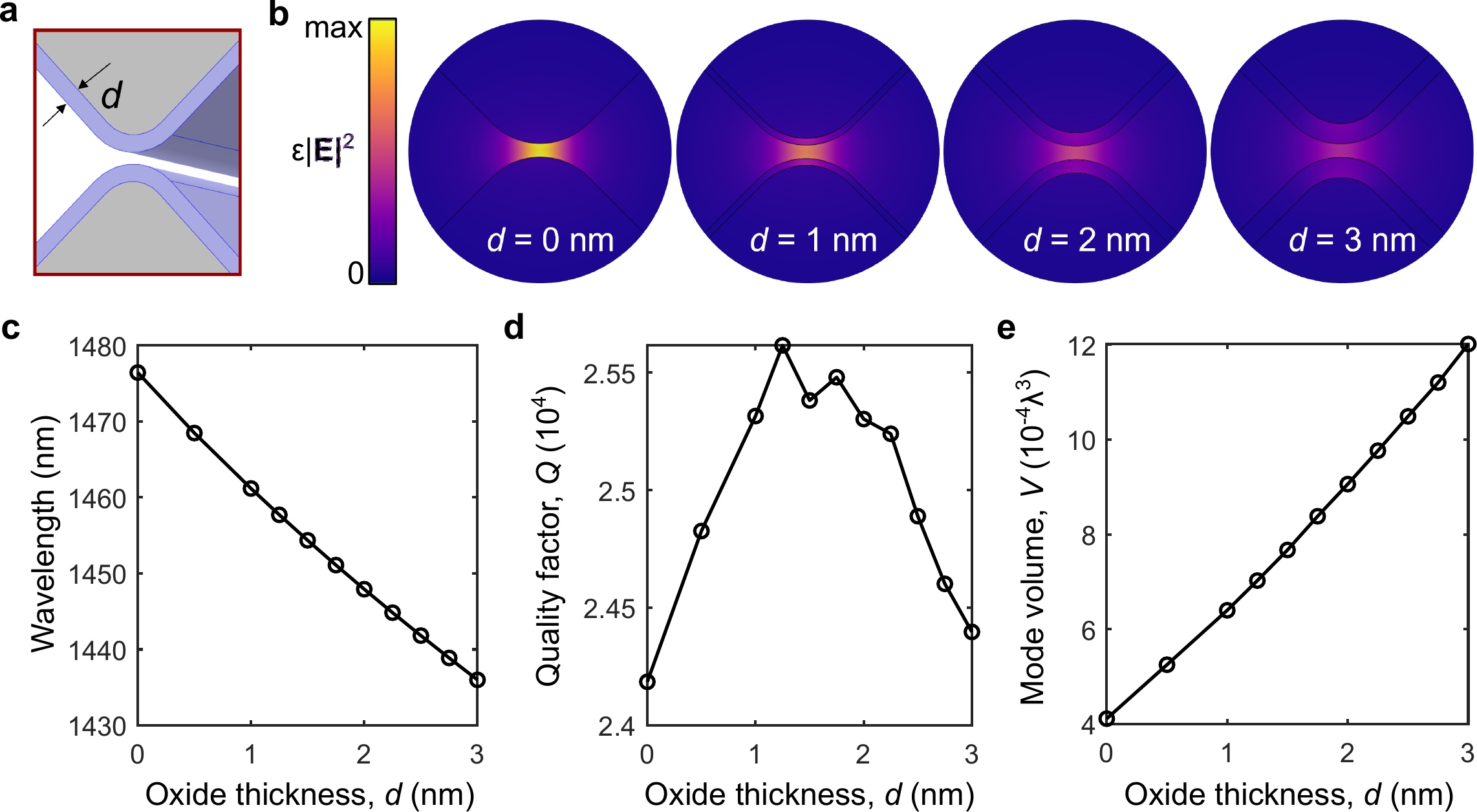}
\caption[]{\textbf{The role of the native oxide on the modal properties of air bowtie nanocavities.} \textbf{a}, Schematic of the conformal native silicon oxide film (blue) in the bowtie region. \textbf{b}, Electric energy density in the air bowtie region of the central unit cell of a cavity with the same geometric parameters as that simulated for Fig. 5 in the main text and varying oxide thickness, $d$. \textbf{c, d, e} The wavelength, quality factor, and mode volume, respectively, of the simulated cavities as a function of $d$.}
\label{fig:Oxide}
\end{figure}

\subsection{S2.2 The role of the native oxide layer}
\label{subsec:oxide}
The simulations in Section~S2.1, which employ an idealized air-bowtie nanobeam structure, do not consider the spontaneously formed native oxide layer in the air-exposed boundaries of the fabricated structures. Through high-resolution scanning transmission electron microscopy (STEM) and electron energy-loss spectroscopy (EELS), we measure the thickness of such native oxide layer (at the bowtie tips) to be between 2 and 2.5 nm (see Subsection~S3.3), in good agreement with previous experimental observations~\cite{nasr2022effect} and ab-initio calculations~\cite{nativeoxide_bohling2016self}. The native oxide layer plays a negligible role for most photonic-crystal cavities, and has therefore generally been ignored in previous works, but the strongly localized fields of bowtie cavities make their modal properties very sensitive to the oxide layer thickness. We illustrate the strong influence of the oxide layer via simulations of the fabricated bowtie cavity for Fig.~5 in the main text with varying native oxide thickness, $d$. Figure~\ref{fig:Oxide}a shows a schematic of the geometry of the simulated bowties, the air gap of which is fixed at $g$ = 2 nm. The effect of the added oxide layer is to reduce the index contrast of the first interface from $\Delta n$ = $n_{\text{Si}}$ - $n_{\text{air}}$ to $\Delta n$ = $n_{\text{oxide}}$ - $n_{\text{air}}$, where the different indices are given by $n_{\text{Si}}$ = 3.48, $n_{\text{oxide}}$ = 1.45 and $n_{\text{air}}$ = 1. Adding the native oxide layer decreases the field intensity in the bowtie, as evidenced by the energy densities shown in Fig.~\ref{fig:Oxide}b. The impact of such field redistribution is negligible on the cavity quality factor, $Q$, but produces a pronounced blue-shift of approximately 14 nm per nm of oxide (Fig.~\ref{fig:Oxide}c) and, more importantly, leads to a more than two-fold increase in the effective mode volume evaluated at the bowtie center, $V$, between a realistic cavity with 2 nm of native oxide and the idealized cavity with no oxide (Fig.~\ref{fig:Oxide}e). We note that the oxide layer is neither included in the collapsed surfaces nor the top and bottom surfaces of the slab in order to limit the number of required mesh elements in the numerical model. 
The considerations of Fig.~\ref{fig:Oxide} were not taken into account in previous works on bowtie nanocavities and reported values of mode volumes from numerical simulations need not only to be regarded with care due to deleterious sources of error such as lightning-rod effects~\cite{Marcus} but also due to the impact of the surface oxide.

\section{S3. Self-assembly of air-bowtie cavities}
\label{sec:selfassembly}

In this section, we discuss the nanofabrication of few-nanometer gaps that can even go down to the atomic scale using conventional lithography and self-assembly.

\begin{figure}[t]
\centering
\includegraphics[width=\textwidth]{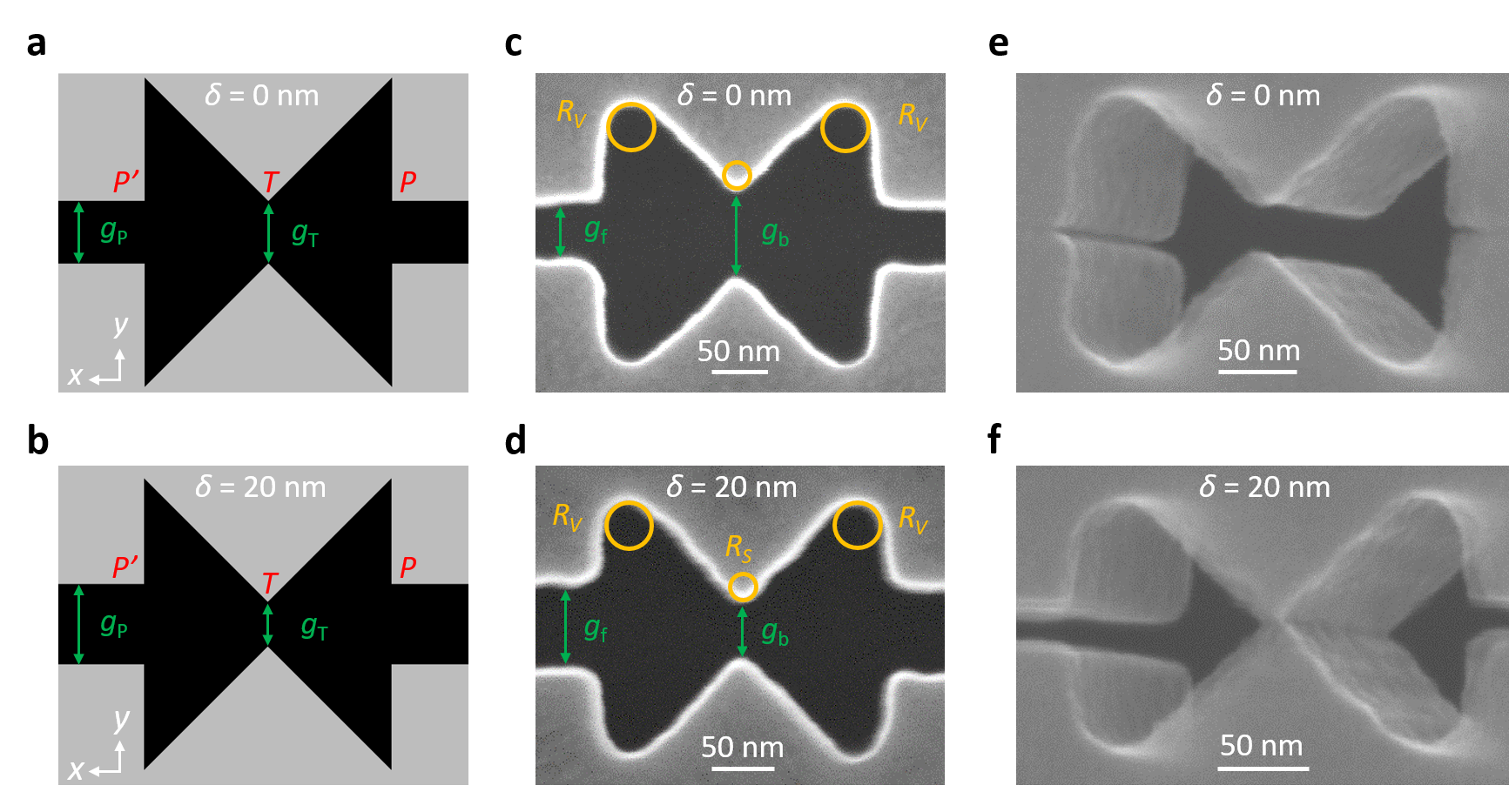}
\caption[]{\textbf{Deterministic fabrication of few-nanometer bowtie widths.} \textbf{a, b}, Schematic of the electron-beam lithography mask for a bowtie unit cell. The exposed region is colored black and the grey region corresponds to silicon features after exposure and reactive ion-etching. The bowtie unit cell comprises two unconnected halves with a nanobeam gap, $g_\text{P}$, and a bowtie gap, $g_\text{T}$. The relative distance between point $T$ and the line $PP'$ is given by the offset $\delta$, which is 0 nm for (\textbf{a}) and 20 nm for (\textbf{b}). \textbf{c, d}, Top-view SEM images of bowtie unit cells fabricated using (\textbf{c}) $\delta$ = 0 nm and (\textbf{d}) $\delta$ = 20 nm, after lithography and plasma etching but before self-assembly. Due to the finite radius of curvature, $R_{\text{S}}$, and the critical dimension loss, $\Delta e$, associated to the fabrication process, $g_\text{P}$ becomes a fabricated gap of width $g_\text{f} = g_\text{P}+\Delta e$ and $g_\text{T}$ becomes $g_\text{b} = g_\text{T} + (\Delta e + R_\text{s})(\sqrt{2}-1)$. \textbf{e, f}, Tilted ($20 ^{\circ}$) SEM images of a bowtie unit cell with (\textbf{e}) $\delta$ = 0 nm and (\textbf{f}) $\delta$ = 20 nm after the self-assembly process. The bowtie with $\delta$ = 0 nm  leads to a pronounced air gap between the tips, and the latter self-assembles into a bowtie with tips in contact and a slot between the parallel surfaces of the nanobeam section.}
\label{fig:relativegap}
\end{figure}

\subsection{S3.1 Deterministic fabrication and size control of air bowties}
\label{sec:Bowtie}

\noindent In the cavity design described in Fig.~\ref{fig:Cavity_design}, all feature sizes except for the few-nanometer air-bowtie widths can be fabricated using conventional lithography and etching. Figures~\ref{fig:relativegap}a and b show schematics of the lithography mask used for the self-assembly of a single bowtie unit cell with a controlled air bowtie width. Relative to the final self-assembled geometry, the mask structure is composed of two unconnected regions that are separated by a gap of width $g_\text{P}$ along the flat parallel boundaries and by a gap of width $g_\text{T}$ between the two bowtie tips. We define the offset, $\delta$, as the relative distance between one of the bowtie tips ($T$) and the line defined by the flat edges ($PP'$), i.e., $\delta =(g_\text{T}-g_\text{P})/2$. For example, $\delta$ = 0 defines a bowtie unit cell where $T$, $P$ and $P'$ are co-linear, and $g_\text{T}$ and $g_\text{P}$ are equal. In contrast, $\delta$ = 20 nm defines a bowtie unit cell where the relative vertical distance between $PP'$ and $T$ is 20 nm, as shown in Figs.~\ref{fig:relativegap}a and b. The reason for having the tips closer to the central axis than the flat edges ($PP'$), i.e., $g_\text{P}>g_\text{T}$, is that the fabrication process rounds all sharp features to finite radii of curvature. We have estimated approximately $R_\text{S}$ = 10 nm for silicon features and $R_\text{V}$ = 20 nm for void features for our fabrication process~\cite{Marcus}. Therefore, the top and bottom tips of the bowties after fabrication are pushed away from each other, as shown in the top-view SEM images of Figs.~\ref{fig:relativegap}c and d, which are taken before underetching.

In addition to changing the offset $\delta$, we also adjust the bowtie dimensions to preserve its $90^{\circ}$ angle, which minimizes shot-filling and fracturing issues during electron-beam lithography. To control the size of the fabricated bowtie, we also take into account the commonly observed process-dependent uniform enlargement, $\Delta e$, of all exposed features. After self-assembly, the resulting bowtie width $g$ is given by
\begin{equation}
\label{eq:fabgap}
    g = 2((\Delta e + R_\text{s})(\sqrt{2}-1)-\delta)
\end{equation}
which determines the largest value of the offset $\delta$ that leads to the formation of a bowtie with an air gap at the tips, i.e., $g>0$, after the self-assembly process. While the solid radius of curvature is well approximated by $R_\text{S}$ = 10 nm, for our fabrication process, the feature growth $\Delta e$ varies from sample to sample with values between 10 nm and 15 nm for the samples in this work. This sets the largest offset that results in an air bowtie to a value between 8.3 nm and 10.4 nm according to Eq.~(\ref{eq:fabgap}). Since the resolution of the electron-beam lithography exposure grid for our fabrication is 1 nm, we expect self-assembled bowtie air gaps to appear for offsets below 11. Since the surface forces mainly originate from the parallel surfaces of the nanobeams, the offset is varied by keeping $g_\text{f}$ fixed at 50 nm while changing $g_\text{b}$. Examples of the resulting air bowties are shown in the tilted-view SEM images of Figs.~\ref{fig:relativegap}e and f for $\delta$ = 0 nm and $\delta$ = 20 nm. The former produces an air bowtie at the unit cell center while the tips are in contact for the latter, which showcases another type of application of the proposed self-assembly in which local protruding regions are used as stoppers, allowing the formation of few-nanometer-wide slot waveguides. In the next section, we demonstrate the deterministic fabrication of few-nanometer gaps by varying the offset $\delta$ between the two extreme cases discussed here.

\begin{figure}[t]
\centering
\includegraphics[width=0.95\textwidth]{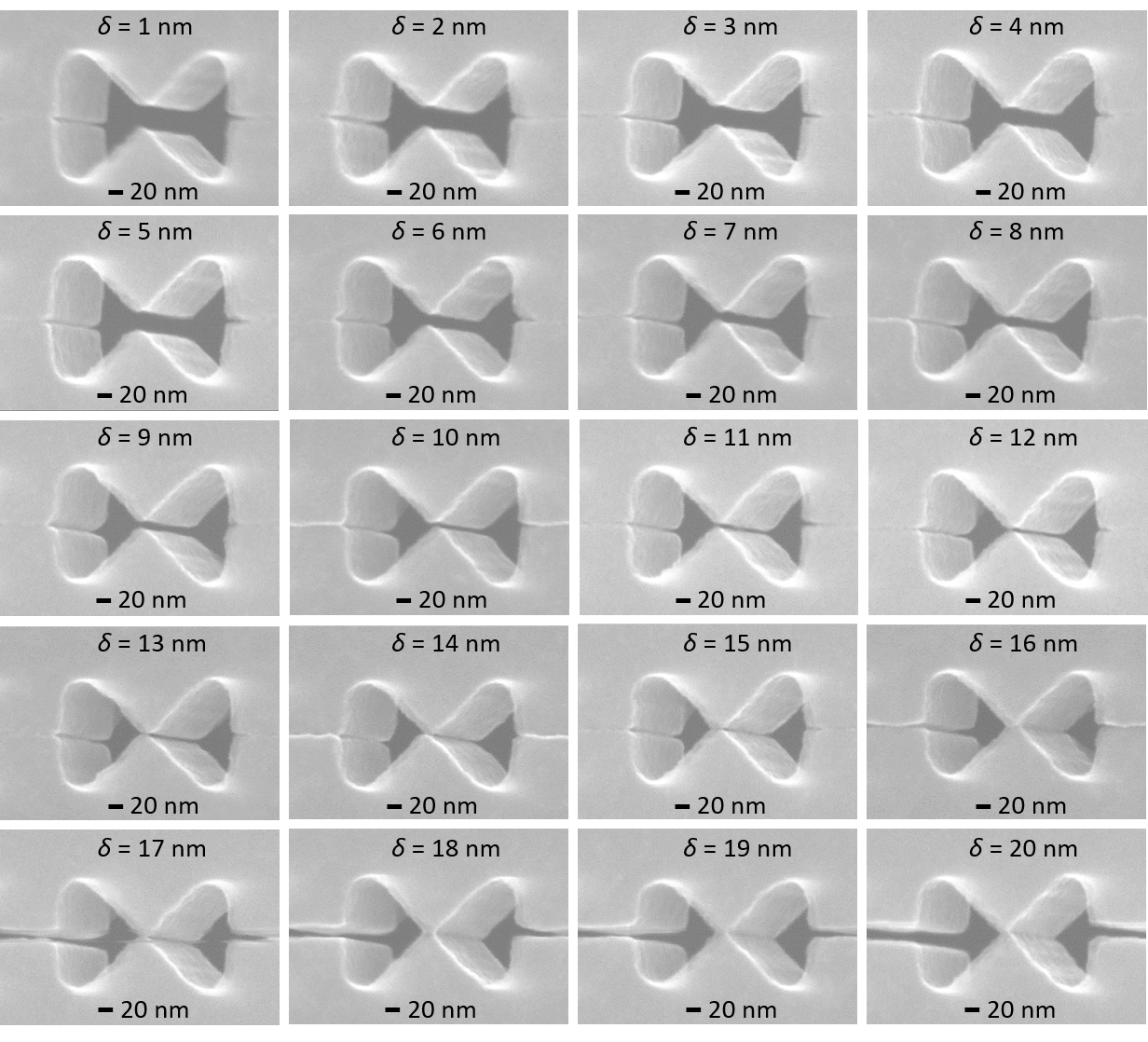}
\caption[]{\textbf{Characterization by SEM of few-nanometer bowtie widths.} Tilted-view ($20 ^{\circ}$) SEM images of an array of self-assembled bowties with offset $\delta$ varying from 1 nm (top-left) to 20 nm (bottom-right).}
\label{fig:Offset}
\end{figure}

\begin{figure}[t]
\centering
\includegraphics[width=0.95\textwidth]{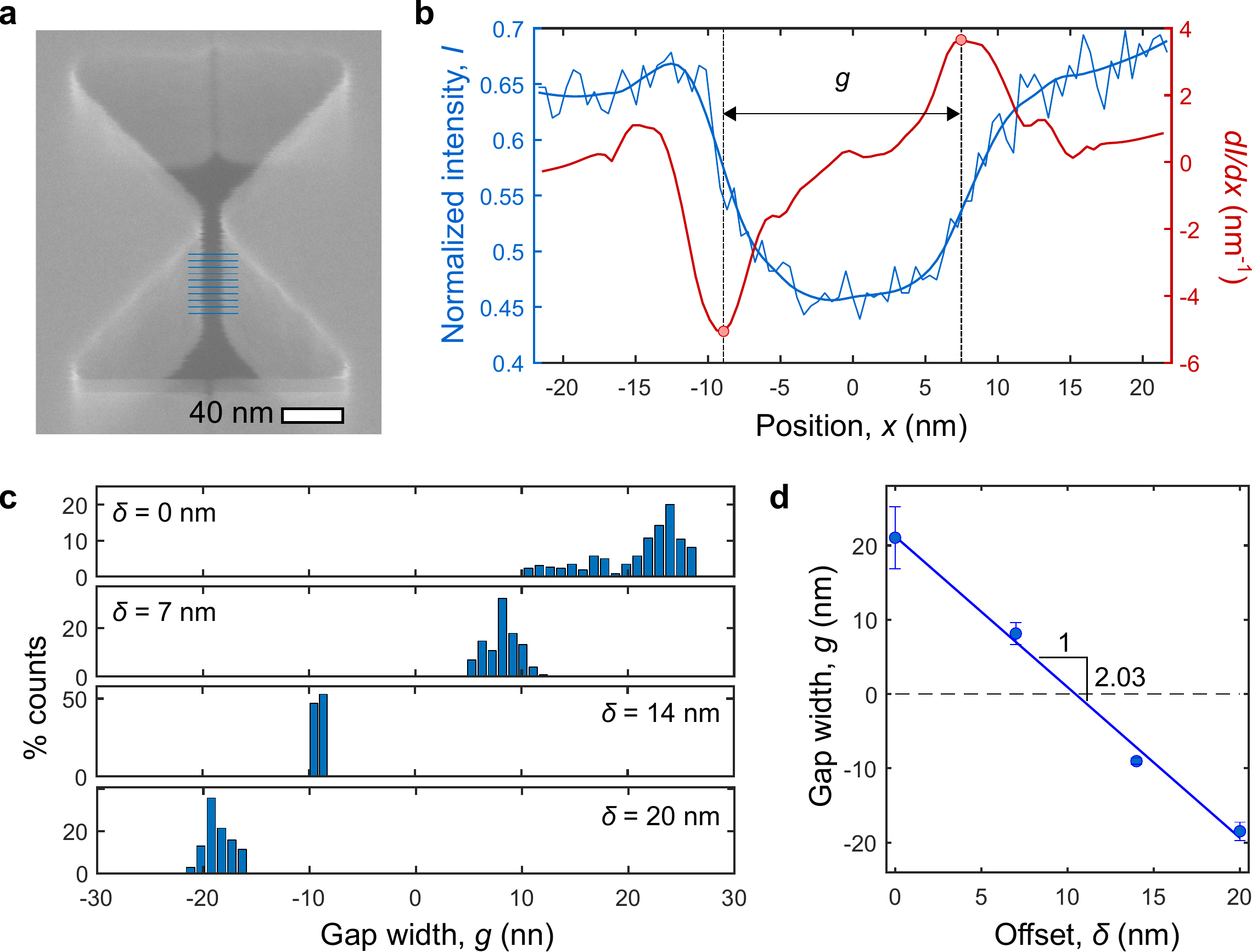}
\caption[]{\textbf{Extracting the relation between the mask offset, $\delta$, and the fabricated bowtie width.} \textbf{a}, Tilted-view ($20 ^{\circ}$) SEM image of a self-assembled air bowtie for $\delta$ = 10 nm. The 10 horizontal lines indicate cuts along which the gap $g$ is extracted via image analysis. \textbf{b}, Normalized intensity along the first cut shown in \textbf{a} and its derivative. The most prominent maxima and minima in the latter are used to extract the gap. \textbf{c} Histograms of the extracted gaps for offsets 0, 7, 14 and 20, from which the average gap are extracted and shown in \textbf{d}, which includes a linear fit (solid blue line).}
\label{fig:SEMextraction}
\end{figure}

\subsection{S3.2 Scanning electron microscope characterization of offset-to-width correspondence}
\label{subsec:SEMcharacterization}

\noindent We fabricate nanobeam cavities by varying $\delta$ from 0 to 20 nm in steps of one nanometer and acquire SEM images on large bowtie subsets to characterize the underlying relation between the offset $\delta$ and the bowtie width $g$ after self-assembly. Figure~\ref{fig:Offset} shows a representative high-resolution tilted SEM image of a bowtie unit cell as $\delta$ is changed from 1 nm (top-left) to 20 nm (bottom-right). First, the bowtie width monotonically narrows until a given offset $\delta_*$, where the bowtie tips touch and there is no void formation at the center. Due to the well-known systematic errors in SEM at the few-nanometer scale as well as bowtie-to-bowtie variations, the exact value of $\delta_*$ cannot be pinpointed exactly, e.g., it is between 13 nm and 15 nm for the set shown in Fig.~\ref{fig:Offset}. As $\delta$ further increases, the bowtie tips start protruding towards the centre from the parallel surfaces and act as stoppers in the directed collapse, generating a slot whose width, also denoted as $g$ for simplicicty, grows monotonically for even larger offsets. Note that, while all the structures in Fig.~\ref{fig:Offset} are fabricated and self-assembled in a single fabrication run, other fabrication runs lead to slightly different offset-to-width correspondence, as is for example the case for the structures reported in Fig. 4c in the main text. In that particular sample, the offsets are limited to $\delta$ = \{0,7,8,9,10,11,12,13,14,20\} nm, which are chosen to ensure covering the regime of few-nanometer air-bowtie cavities while including two cases in which the structure exhibits wide bowties ($\delta$ = 0 nm) and wide slots ($\delta$ = 20 nm) to help find the precise offset-to-width relation without being limited by the SEM resolution or artifacts. We image all the bowtie unit cells in the cavity region of several nanobeams fabricated with offsets 0, 7, 14 and 20 and extract the gap width in 10 positions across the device layer thickness for each image, as exemplified in Fig.~\ref{fig:SEMextraction}a. At each position, $g$ is extracted via edge detection using the maximum derivative points in a smoothed version of the SEM image intensity (Fig.~\ref{fig:SEMextraction}b). Figure~\ref{fig:SEMextraction}c depicts the histogram obtained for the 4 used offsets, while Fig.~\ref{fig:SEMextraction}d depicts a linear fit to the average gaps that we use to estimate the effective widths at all other values of $\delta$, notably the bowtie widths of 5.3 nm, 3.3 nm and 1.3 nm used for $\delta$ = 8, 9 and 10 in Fig. 4c and d in the main text. We highlight that the extracted slope of 2.03 is in excellent agreement with the expected value of 2 and that the curve intersects around $\delta_*$ = 11 nm, in good agreement with the estimate done via Eq. (\ref{eq:fabgap}).

\subsection{S3.3 Scanning transmission electron microscope characterization of bowtie widths}
\label{subsec:STEMcharacterization}

\begin{figure*}[t]
\centering
 \includegraphics[width=0.75\textwidth]{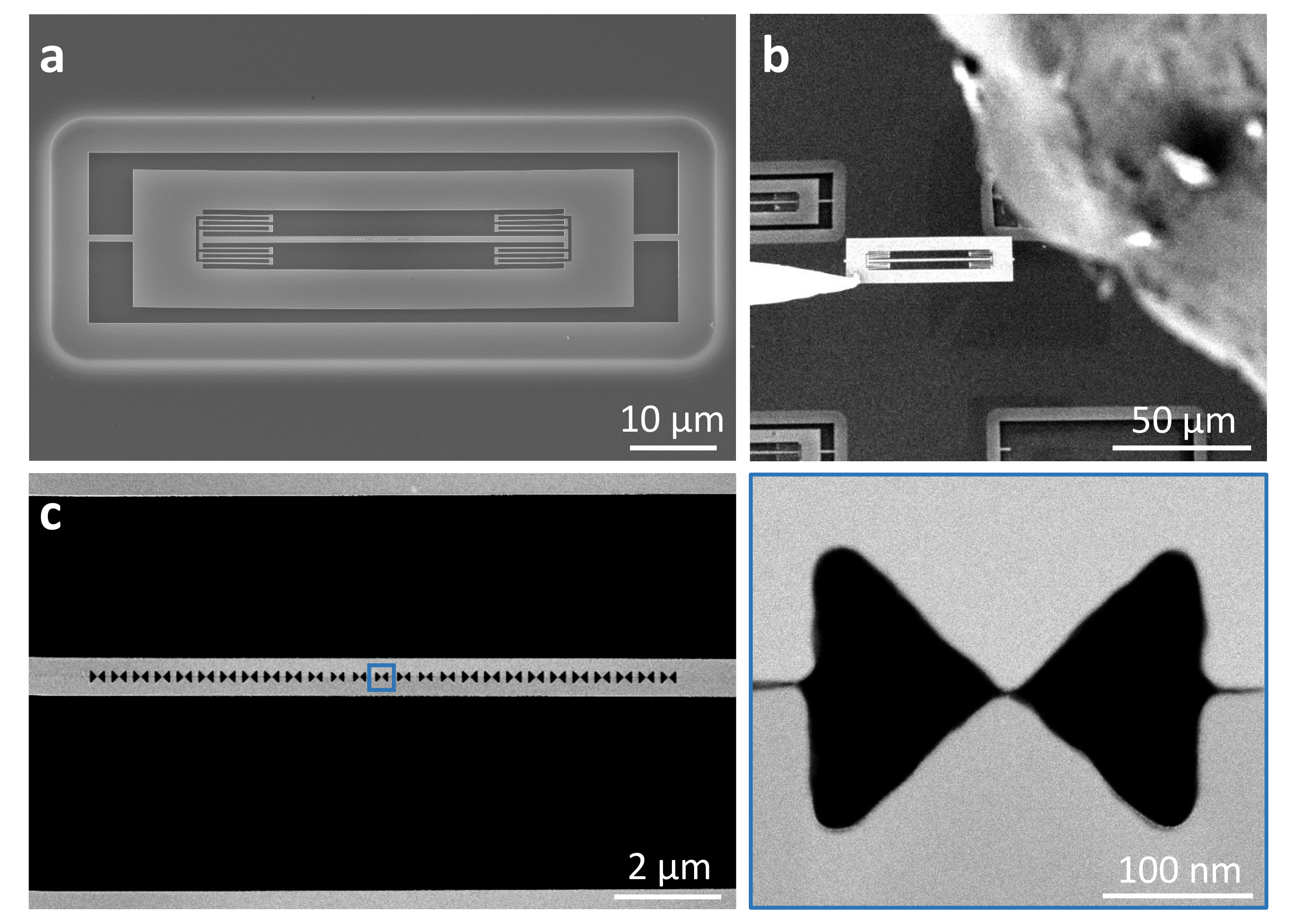}
    \caption{\textbf{High-resolution STEM imaging of self-assembled nanobeam cavities.} \textbf{a}, Tilted ($20 ^{\circ}$) SEM image of a self-assembled nanobeam cavity with the surrounding frame designed for FIB-assisted lift-off. \textbf{b}, Transfer of a nanobeam to a STEM grid using a micromanipulator tip. \textbf{c}, Top-view STEM image of a self-assembled nanobeam cavity and a zoom-in into the central unit cell (blue box).
    }
    \label{fig:FIB}
\end{figure*}

 \begin{figure}[t]
\centering
\includegraphics[width=\textwidth]{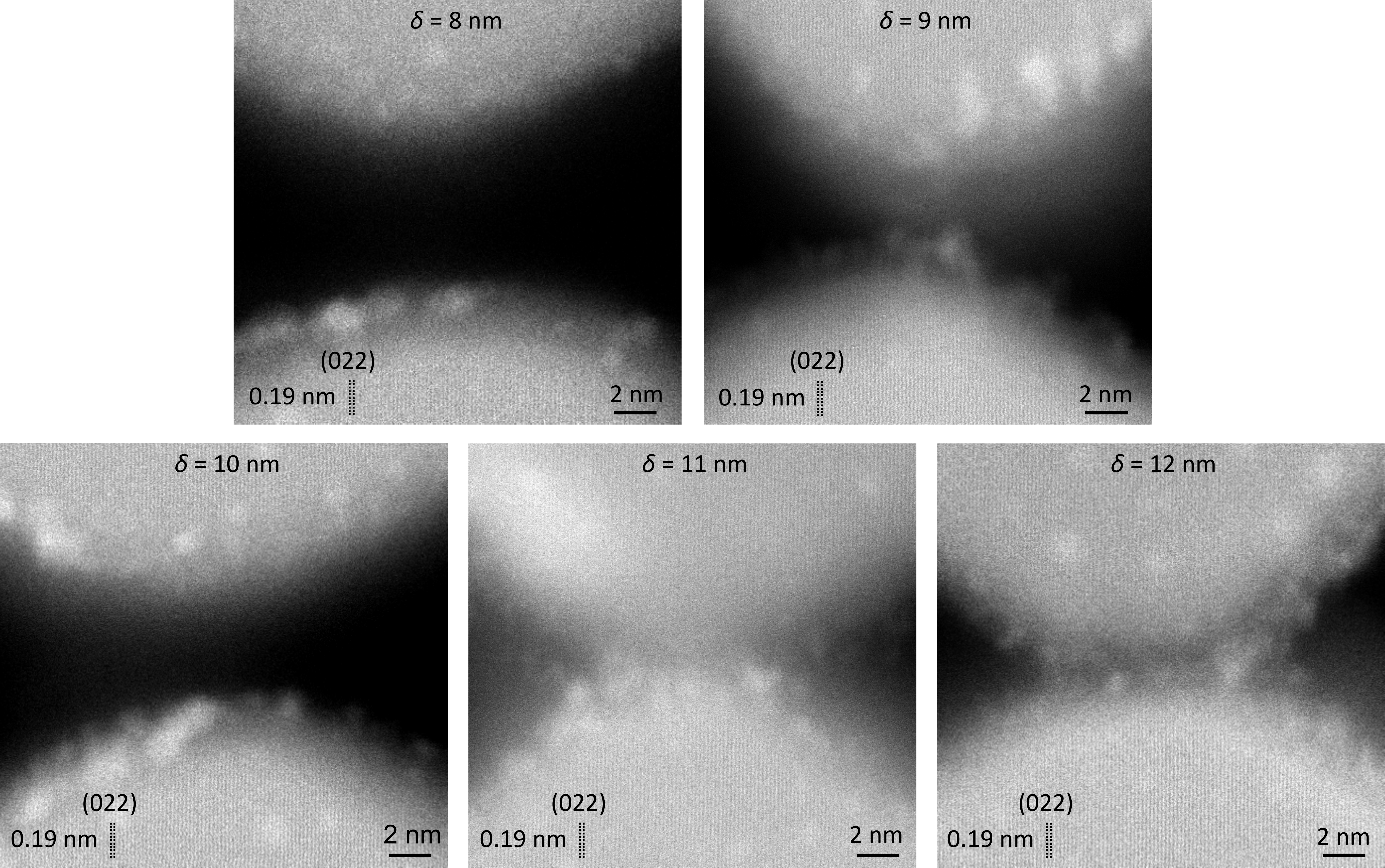}
\caption[]{\textbf{Characterization by STEM of atomic-scale bowtie widths.} Annular dark-field STEM images of the bowtie tips of the central unit cells in nanobeam cavities fabricated with $\delta$ from 8 nm to 12 nm. The (022) crystal planes of silicon are observed and indicated in the images.}
\label{fig:TEM}
\end{figure}

\noindent To precisely characterize the bowtie width for values of $\delta$ around $\delta_*$ and the native oxide layer, we fabricate self-assembled cavities for high-resolution STEM imaging. Since STEM requires thin layers of material to transmit electrons for analysis, individual nanobeam cavities are cut out from the wafer chip using a focused ion beam (FIB) and transferred to STEM-compatible grids (see Methods for details on the lift-off process for STEM imaging). In the sample dedicated to STEM, we employ an anchoring system that assists in cutting and lifting off the cavities from the chip (see Fig.~\ref{fig:FIB}a). We select $5$ devices from the sample with $\delta$ = 8, 9, 10, 11, and 12 nm and transfer them to the TEM grids using a motorized micro-manipulator as shown in Fig.~\ref{fig:FIB}b. Figure~\ref{fig:FIB}c shows a top-view STEM image of a self-assembled nanobeam cavity and its central bowtie unit cell. High-resolution STEM images of the central bowtie region for $\delta$ of 8 to 12 nm are shown in Fig.~\ref{fig:TEM}.

The signal intensity in STEM depends on composition, density, and thickness and it decreases gradually when transitioning from the crystalline silicon to the amorphous oxide and reducing to a background signal at the void region. Given the spatial resolution of approximately 0.1 nm of our STEM and the observations of the sidewall tilt and roughness discussed in the main text, we attribute the intensity drop to the bowtie geometry and composition, i.e., the thinner the probed thickness, the dimmer the intensity. We extract the thickness of the native oxide layer by considering its edges to be defined by the position where the crystalline lattice is no longer visible on the bottom half bowtie side and where the rate of change in intensity is maximum on the void side. Based on where the (022) lattice signal extends in the images, a 2-2.5 nm thick amorphous layer is measured at the edge of the structures, which we attribute to being the native silicon oxide layer. The exact measurement is complicated due to the dependence of signal intensity with probed thickness. As expected from the observed negative correlation between offset and bowtie width, a pronounced air gap (void region) is observed for $\delta$ = 8 nm. At offsets 11 and 12, the bowtie tips are in contact at the native oxide layers, therefore, the void region disappears. In between, the progressive change in the bowtie width is more complex than one would expect from the design rules described in Section~S3.1 due to a combination of the roughness of the bowtie tips and the bowtie shape in the vertical direction.

\section{S4. Optical spectroscopy of air-bowtie nanocavities}

\subsection{S4.1 Far-field resonant scattering measurements}

\begin{figure}[t]
\centering
\includegraphics[width=0.85\textwidth]{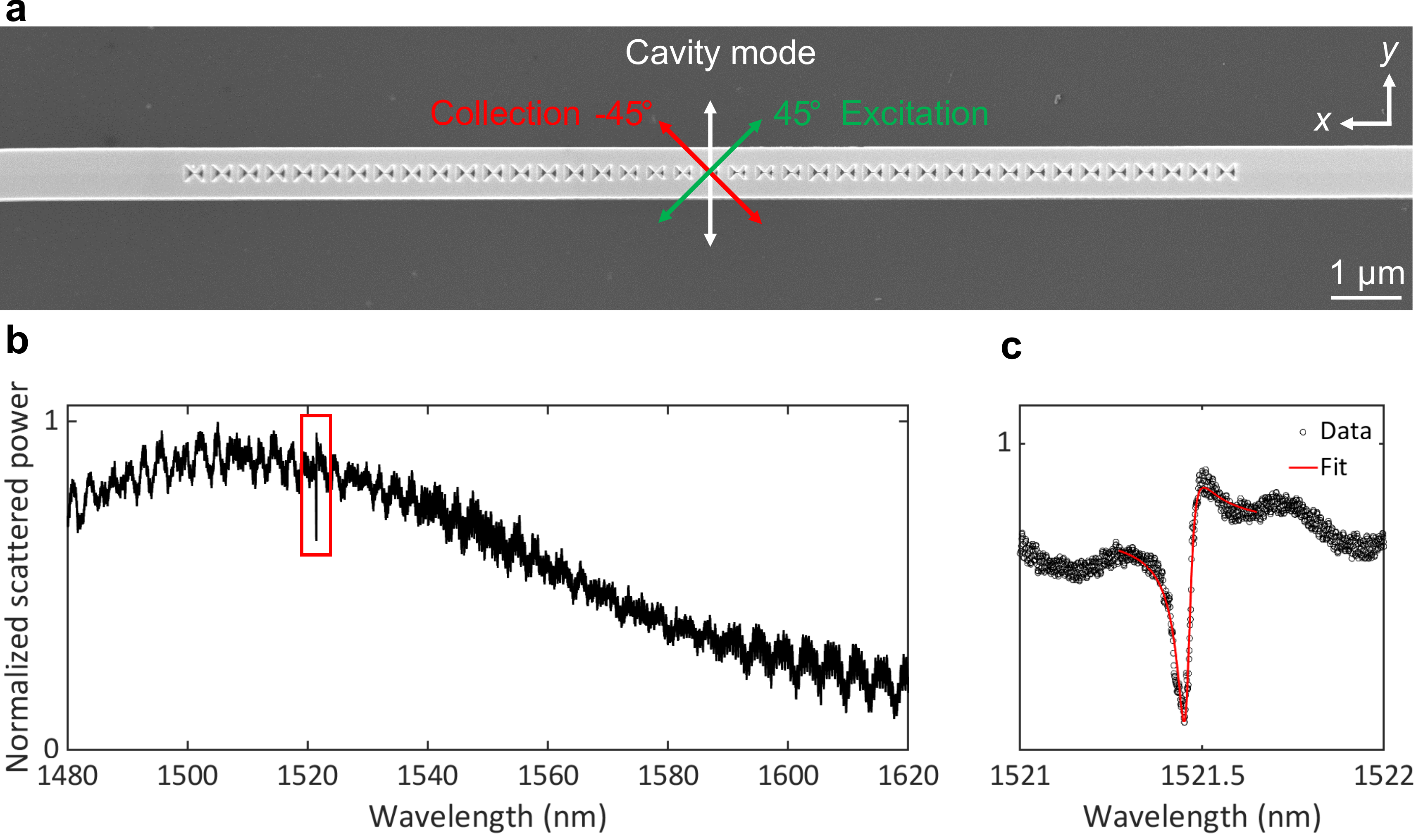}
\caption[]{\textbf{Cross-polarized resonant-scattering spectroscopy.} \textbf{a}, Tilted-view ($20 ^{\circ}$) SEM image of a self-assembled nanobeam cavity showing the polarization of excitation (green arrow), detection (red arrow) and the cavity mode (white arrow). \textbf{b}, Normalized measured scattered power with excitation and collection at the cavity center. The red box highlights the cavity resonance, also shown in \textbf{c}. A fit to a Fano lineshape is overlayed in red, from which we extract the resonance wavelength and quality factor.}
\label{fig:Fanofit}
\end{figure}

\noindent As detailed in the Methods section, we perform optical spectroscopy of the self-assembled nanocavities using far-field resonant scattering measurements. The probed cavity mode has a polarization in the far field that is mainly along the $y$-axis as indicated with the white arrow in the SEM image of Fig.~\ref{fig:Fanofit}a. We couple light into the cavity by exciting at normal incidence using linearly polarized light with a polarization of $45^{\circ}$ relative to the cavity mode polarization and collect light at $90^{\circ}$ relative to the excitation polarization. The excitation and collection with cross-polarization optimizes the cavity-to-background coupling efficiency. The cavity resonances appear as Fano resonances due to the interference between the high-$Q$ in-plane cavity resonance and the low-$Q$ out-of-plane resonance formed by the silicon membrane and the handle layer~\cite{Marcus}. The measured spectrum shown in Fig.~\ref{fig:Fanofit}b corresponds to one of the self-assembled cavities used for Fig.~4a in the main text ($\delta$ = 8 nm). After fitting a Fano lineshape to the resonant feature (Fig.~\ref{fig:Fanofit}c), we extract a resonant wavelength of $\lambda$ = 1521.5 nm and a quality factor of approximately $3.9\times 10^4$.

\begin{figure}[t]
\centering
\includegraphics[width=1\textwidth]{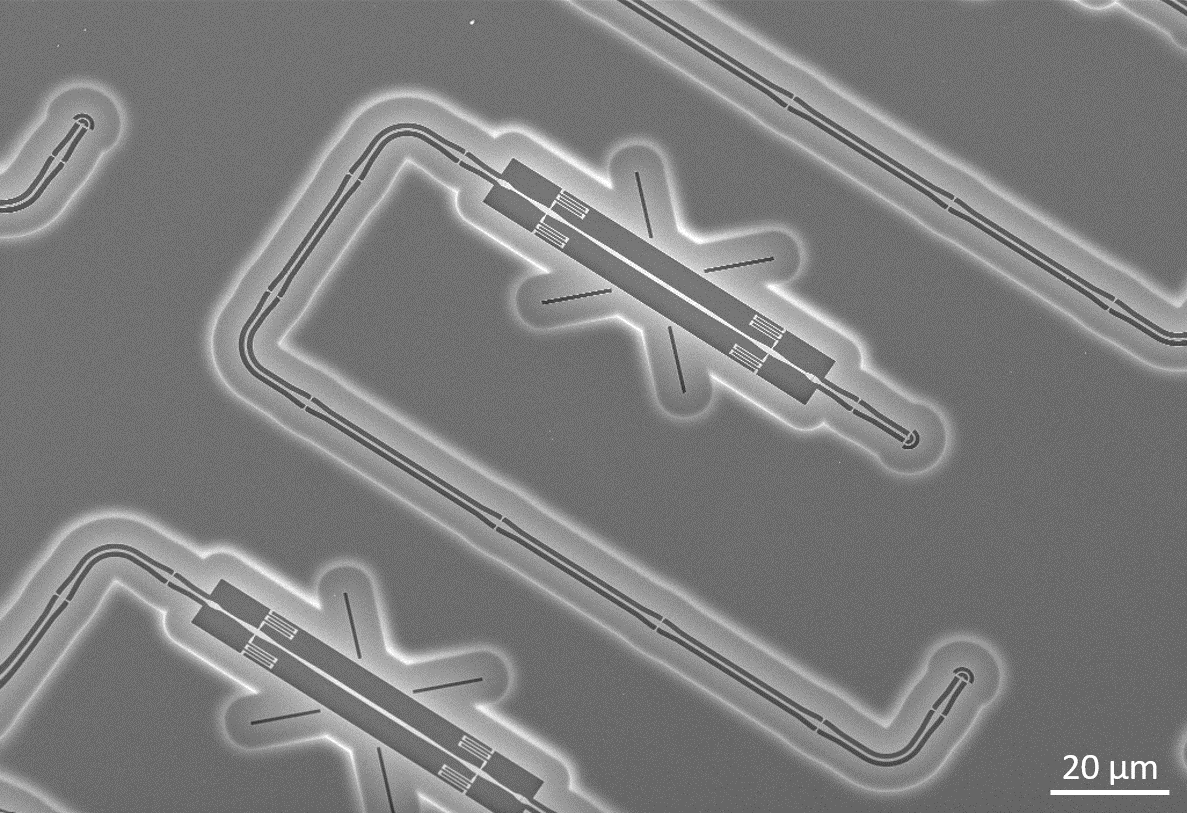}
\caption[]{\textbf{Suspended photonic circuits with self-assembled nanobeam cavities.} Tilted ($20 ^{\circ}$) SEM image of a photonic circuit to characterize a self-assembled nanobeam cavity via in-plane transmission measurements.}
\label{fig:waveguidecoupledcavity}
\end{figure}

\subsection{S4.2 In-line transmission measurements of self-assembled nanobeam cavities}

\begin{figure}[t]
\centering
\includegraphics[width=0.9\textwidth]{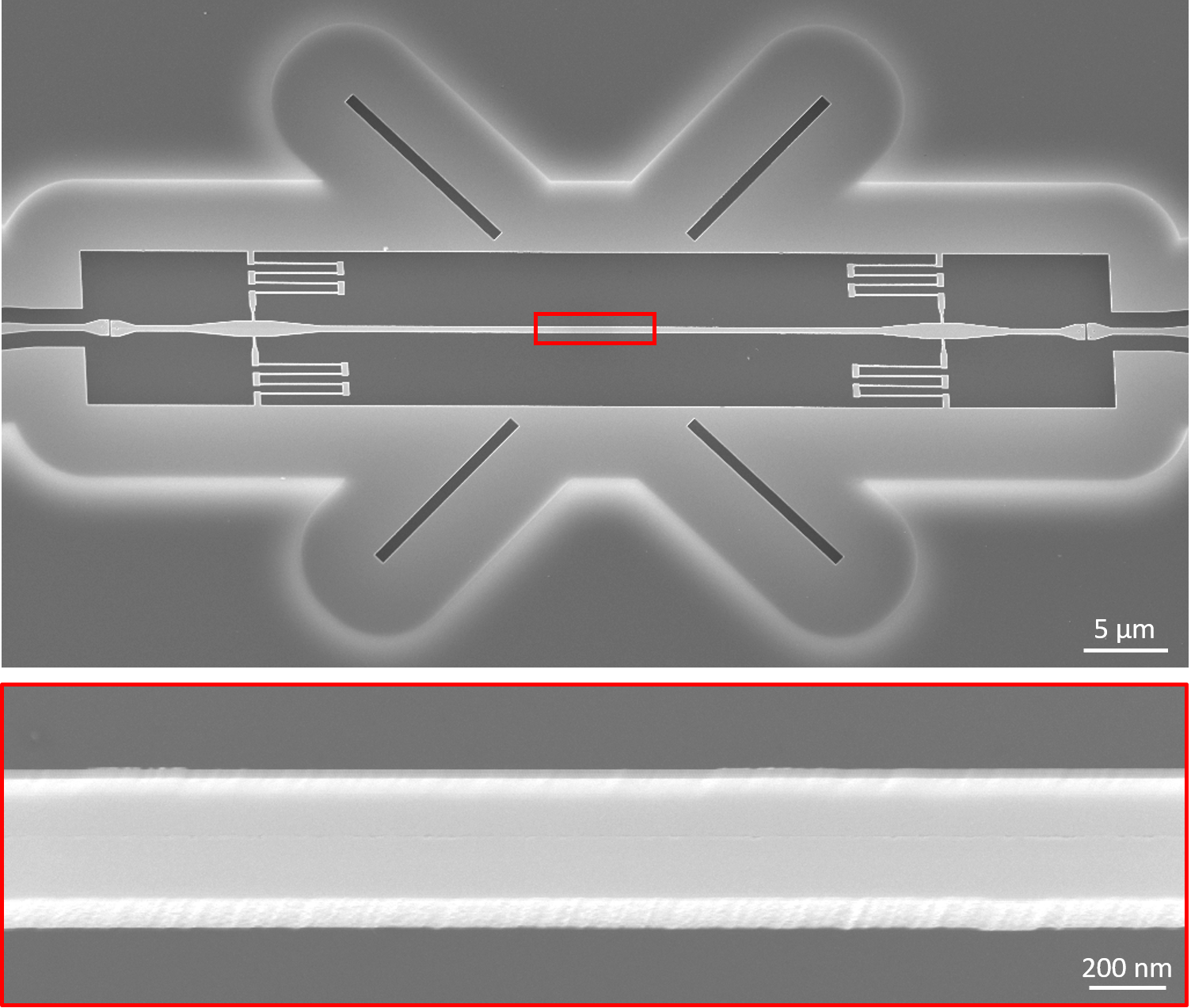}
\caption[]{\textbf{Suspended self-assembled nanobeam waveguides.} Tilted ($20 ^{\circ}$) SEM image of a self-assembled nanobeam waveguide, the transmission of which is used to normalize the transmission measurements on self-assembled cavities. The red box is a zoom-in of the central part of the self-assembled waveguide.}
\label{fig:selfassembledwaveguide}
\end{figure}

\begin{figure}[t]
\centering
\includegraphics[width=0.9\textwidth]{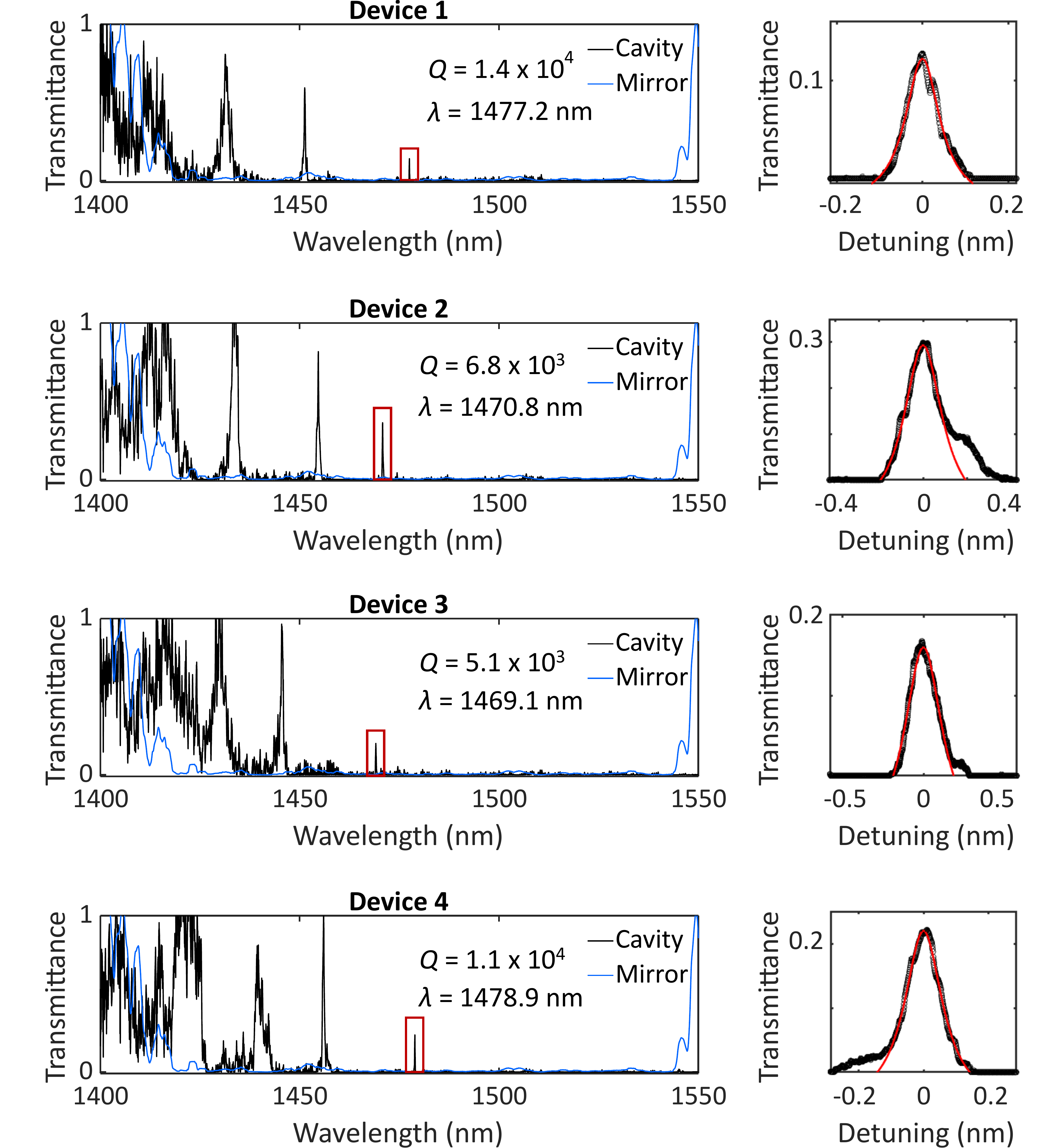}
\caption[]{\textbf{Optical spectroscopy of nominally identical waveguide-coupled self-assembled nanobeam cavities.} Transmittance spectra of 4 nominally identical bowtie cavities (Cavity) and bowtie waveguides (Mirror) with an approximate bowtie width of 2 nm. On each panel, a red box highlights the fundamental cavity resonance, a Lorentzian fit to which is shown separately. The extracted loaded quality factor and resonant wavelength are indicated.}
\label{fig:Clones}
\end{figure}

\noindent We perform optical spectroscopy of waveguide-coupled self-assembled nanocavities using in-line transmission measurements, as detailed in the Methods section. Figure~\ref{fig:waveguidecoupledcavity} shows a tilted SEM image of an entire waveguide-coupled self-assembled nanobeam cavity device, including the two circular grating couplers used for cross-polarized and spatially resolved excitation/collection. As discussed in the main text, the spectra obtained on such cavities are normalized to that measured on a self-assembled waveguide, a characteristic SEM image of which is shown in Fig.~\ref{fig:selfassembledwaveguide}. The inset shows that the quality of the collapse is such that the interface is hardly visible. However, the observed line-edge roughness on the sidewalls also evidences that the proposed normalization is likely more appropriate than normalizing with a conventional suspended waveguide. To demonstrate the robustness of our self-assembly method, we fabricate 4 nominally identical circuits for the waveguide-coupled self-assembled nanocavity shown in the main text Fig.~5. Fig.~\ref{fig:Clones} shows the normalized transmission measurements for the set of copies, including a measurement of a nanobeam waveguide of the same length and with the unit cell corresponding to that of the cavity mirrors. For all the cavities, we observe a reduction of the quality factor and the on-resonance transmission relative to the simulated values, respectively $Q$ = $4\times10^{4}$ and $T_\text{o}$ = 0.96. We attribute the former drop to the effect of the self-assembled interface in the collapsed regions and to the sidewall roughness around the air bowties where the cavity field is most intense. That same drop partly explains the drop in $T_\text{o}$, which also occurs due to the differential bowing of the structure observed for the nanobeam cavities and the nanobeam waveguide, i.e., the normalization is not perfect. In addition, both the measured $Q$ and $T_\text{o}$ fluctuate considerably, indicating a subtle interplay between the disorder-induced losses in the structure plane and the out-of-plane direction.

\clearpage
